\documentclass[longauth]{aa}

\usepackage{graphicx}
\usepackage{yfonts}
\usepackage{xcolor}
\usepackage{txfonts}
\usepackage{subfigure}
\usepackage{float}
\usepackage{natbib,twoopt}
\usepackage{xspace}
\usepackage{balance}
\usepackage{upgreek}
\usepackage{multirow}
\usepackage[normalem]{ulem}
\usepackage{soul}
\usepackage[colorlinks=true, citecolor=blue, linkcolor=blue]{hyperref}
\usepackage[normalem]{ulem}
\usepackage{siunitx}
\usepackage[flushleft]{threeparttable}
\usepackage[nameinlink, capitalise]{cleveref}
\usepackage{orcidlink}
\usepackage{amstext}

\def\apjl{ApJ Lett.}
\def\apjs{ApJ Suppl.}

\def\aap{A\&A}

\bibpunct{(}{)}{;}{a}{}{,} 
\DeclareGraphicsExtensions{.eps,.ps,.pdf}
\begin{document}
\title{AMICO galaxy clusters in KiDS-DR3: measuring the splashback radius from weak gravitational lensing}
\author
{Carlo Giocoli\orcidlink{0000-0002-9590-7961}\inst{\ref{1},\ref{2}}, Lorenzo Palmucci\inst{\ref{3}}, Giorgio F. Lesci\inst{\ref{3},\ref{1}}, Lauro Moscardini\inst{\ref{3},\ref{1},\ref{2}}, Giulia Despali\inst{\ref{3},\ref{1},\ref{2}}, Federico Marulli\inst{\ref{3},\ref{1},\ref{2}}, 
Matteo Maturi \inst{\ref{5},\ref{6}},
Mario Radovich\inst{\ref{4}}, Mauro Sereno\inst{\ref{1},\ref{2}},
Sandro Bardelli\inst{\ref{1}}, Gianluca Castignani\inst{\ref{1}},
Giovanni Covone\inst{\ref{nove},\ref{dieci},\ref{undici}}, 
Lorenzo Ingoglia\inst{\ref{3}},
Massimiliano Romanello\inst{\ref{3},\ref{1}}, Mauro Roncarelli\inst{\ref{1},\ref{2}}, Emanuella Puddu\inst{\ref{dieci}}} 
\offprints{\\ \email{carlo.giocoli@inaf.it}}

\institute{
INAF-Osservatorio di Astrofisica e Scienza dello Spazio di Bologna, Via Piero Gobetti 93/3, 40129 Bologna, Italy
\label{1} \and 
INFN-Sezione di Bologna, Viale Berti Pichat 6/2, 40127 Bologna, Italy \label{2} \and
Dipartimento di Fisica e Astronomia "Augusto Righi" - Alma Mater Studiorum Universit\`{a} di Bologna, via Piero Gobetti 93/2, 40129 Bologna, Italy
\label{3} \and 
INAF - Osservatorio Astronomico di Padova, vicolo dell'Osservatorio 5, I-35122 Padova, Italy \label{4} \and 
Zentrum f\"ur Astronomie, Universit\"at Heidelberg, Philosophenweg 12, D-69120 Heidelberg, Germany \label{5} \and
\label{6} ITP, Universit\"at Heidelberg, Philosophenweg 16, D-69120 Heidelberg, Germany \and
\label{nove} Dipartimento di Fisica "E. Pancini", Universit\'a di Napoli Federico II, C.U. di Monte Sant'Angelo, via Cintia, I-80126 Napoli, Italy \and
\label{dieci} INAF - Osservatorio Astronomico di Capodimonte, Salita Moiariello 16, I-80131, Napoli, Italy  \and
\label{undici} INFN - Sezione di Napoli, via Cintia, I-80126, Napoli, Italy 
}

\date{Received --; accepted --}

\abstract {Weak gravitational lensing offers a powerful method to investigate the projected matter density distribution within galaxy clusters, granting crucial insights into the broader landscape of dark matter on cluster scales.}
{In this study, we make use of the large photometric galaxy cluster data set derived from the publicly available Third Data Release of the Kilo-Degree Survey, along with the associated shear signal. Our primary objective is to model the peculiar sharp transition in the cluster profile slope, that is what is commonly referred to as the splashback radius.  
The data set under scrutiny includes 6962 galaxy clusters, selected by \texttt{AMICO} -- an optimised detection algorithm of galaxy clusters -- on the KiDS-DR3 data, in the redshift range of $0.1 \le z \le 0.6$, all observed at a signal-to-noise ratio greater than 3.5.}
{Employing a comprehensive Bayesian analysis, we model the stacked excess surface mass density distribution of the clusters. We adopt a model from recent results on numerical simulations that capture the dynamics of both orbiting and infalling materials, separated by the region where the density profile slope undergoes a pronounced deepening.}
 {We find that the adopted profile successfully characterizes the cluster masses, consistent with previous works, and models the deepening of the slope of the density profiles measured with weak-lensing data up to the outskirts.
Moreover, we measure the splashback radius of galaxy clusters and show that its value is close to the radius within which the enclosed overdensity is 200 times the mean matter density of the Universe, while theoretical models predict a larger value consistent with a low accretion rate. This points to a potential bias of optically selected clusters preferentially characterized by a high density at small scales compared to a pure mass-selected cluster sample.}
{}
    
\authorrunning{Giocoli et al.}
\titlerunning{AMICO galaxy clusters in KiDS-DR3: splashback radius from weak-lensing}
\maketitle

\section{Introduction}

In the standard structure formation scenario, dark matter haloes represent the building blocks of galaxy assembly processes \citep{white78,lacey93,tormen98a}. 
Systems assemble via repeated merging events, and haloes hosting galaxy clusters are the last forming structures at the top of this hierarchical pyramid \citep{springel01b,vandenbosch02,giocoli07}.  
In the last three decades, numerical simulations have clarified this picture, allowing us to define the sizes and boundaries of dark matter haloes hosting hundreds or thousands of galaxies \citep{springel10,bonafede11,cui18}.

Following the spherical collapse model, the halo mass $M_{\rm vir}$ is defined as the mass within the radius enclosing the virial overdensity $\Delta_{\rm vir}$ \citep{press74,bardeen86,sheth01b}. 
Other common choices are the masses corresponding to an overdensity of 200 times the Universe's critical or mean background density ($M_{\rm 200c}$ and $M_{\rm 200m}$ respectively).
These translate into a different mass definition for the same object, and generally $M_{\rm 200c}\leq M_{\rm vir} \leq M_{\rm 200m}$ \citep{angulo09,tinker08,crocce10,angulo12,despali16}. The dark matter distribution around clusters thus represents crucial cosmological information. The halo concentration, expressed as the ratio between $r_s$ -- the radius at which the logarithmic derivative of the density profile is equal to -2 -- and the halo radius, is, on average, a decreasing function of the halo mass \citep{bullock01a,giocoli12b}. 
In the hierarchical picture of structure formation, $r_s$ settles to a constant value after a rapid formation and accretion phase, while the halo radius evolves due to merging events \citep{vandenbosch02,wechsler02,giocoli08b} and cosmological {\em pseudo-evolution} depending on the considered overdensity mass \citep{diemer13}.

Recently, a different definition of the halo boundary has been proposed \citep{diemer14,adhikari14}, related to the transition region between the orbiting and infalling mass components. 
It corresponds to the distance at which satellite galaxies and matter, after their first apocentric passage, splash back toward the halo centre: accreting matter is in the process of reversing direction to fall back into the halo under the influence of gravitational forces.
The term splashback radius is used to describe this distance and, in practice, it is close to the turnaround radius of matter that previously fell into the halo when following the nonlinear evolution of collapsing spherical shells \citep{gunn72,fillmore84,bardeen86}. 

It is worth underlining that the density profiles of gravitationally bound structures mirror their hierarchical growth along the cosmic time as a consequence of repeated merging events. In this respect, at a fixed dark matter halo mass, the location of the splashback radius reflects the recent accretion history via the accretion rate. In particular, the splashback radius is very sensitive to the recent accretion rate over the past crossing time and to the concentration parameter \citep{more15,shin23}. 

Many recent works modelled the splashback radius and its connection to the density profiles in numerical simulations. 
Using phase space distributions of dark matter particles, \citet{garcia23} have shown that orbiting particles (with early accretion time) have low mean radial velocities, whereas infalling particles (with late accretion time) have large negative radial velocities, and the transition is expected when particles reach their first apocentric passage. 
They also show that the corresponding mass function of orbiting particles can be described by the standard \citet{press74} formula but with a peak height dependent threshold barrier parameter for collapse $\delta_{sc}$.
\citet{pizzardo24} have used the radial velocity profile to measure the turnaround, infall and radius of minimum radial velocity on a large sample of clusters from the Illustris TNG-300 simulation \citep{pillepich18,nelson18,springel18}. 
They have also given two other radius definitions, one from setting the ratio between the average velocity dispersion and radial velocity profiles equal to $-1$,  termed as $R_{\sigma_v}$, and the other radius from the satellite galaxy spacial profile that they call the splashback radius $R^{n_g}_{sp}$. 
In their work, they underline that within the confidence interval of 1 $\sigma$, the infall radius $R_{\sigma_v}$ and $R^{n_g}_{sp}$ coincide.
\citet{xhakaj20} have studied the characterization of the splashback radius in simulations using different methods to shed more light on the observational analyses. 
They underline that by using satellites, the recovered three-dimensional profile traces the depth of the halo gravitational potential well, and the splashback radius definition is in agreement with what has been measured by their adopted halo finder code SPARTA \citep{diemer17,diemer20}. 
In addition, they stress that the best method to define the splashback radius observationally is through weak-gravitational lensing since it does not rely on galaxies to trace the host halo potential well and does not depend on the corresponding satellite dynamical friction effects that can create biases \citep{umetsu17}. 
It is also interesting to notice that high and intermediate accretion rate clusters manifest a secondary caustic in the density profile that can cause biases in constraining the splashback radius. 
In lensing, the second caustic is washed out, which possibly gives a better definition of the splashback radius. 
Recently \citet{towler24}, using the Flamingo simulation data set \citep{shaye23}, looked for a possible correlation between the splashback radius and gas properties, finding that the location of the minimum in the gas gradient is not directly related to the position of the splashback radius. 

Beyond the splashback radius, the density profile of the halo steeply declines. 
Therefore, it serves as a crucial parameter in characterising the spatial extent of dark matter haloes and provides insights into the dynamic processes of matter accretion along the cosmic web. This has been explored and validated through both numerical simulations \citep{diemer14,more15,xhakaj20,pizzardo24} and observational data \citep{baxter17,chang18,murata20,adhikari21}. 
It is worth highlighting that the location could also represent a cosmological test which can be used to look for possible signatures beyond standard $\mathrm{\Lambda}$-cold dark matter ($\mathrm{\Lambda}$CDM) model \citep{adhikari18,contigiani19a,despali20,despali22}. 
At a fixed accretion rate, the splashback radius depends on the dark energy equation of state parameter $w$; being related to the expansion history of the Universe, it tends to be larger for lower $w$. \citet{adhikari18} have shown that the location of the splashback radius depends on the gravity model too. 

Using galaxy cluster data, different authors have explored the galaxy cross-correlation \citep{chang18,zurcher19,shin19,murata20,shin21,rana23,contigiani23}, that probes the satellite galaxy distribution to model the projected matter density profiles and constrain the splashback radii. 
However, it is worth noticing that in these methods, the radius at which the density profile exhibits a sharp steepening of the slope depends on the selection in galaxy magnitudes and colours \citep{murata20}. Red and luminous galaxies are typically more centrally concentrated than the blue and faint ones, resulting in different splashback radii for the two groups, even if consistent within measurement uncertainties \citep{murata20}. 

Following \citet{umetsu17,contigiani19a,shin21,rana23},  we base our analysis on weak gravitational lensing of optically selected clusters to model their projected matter density distribution. We believe this method provides an unbiased definition of the radius at which the density profile slope steepens, free from possible systematic uncertainties related to the satellite galaxy selection. Nonetheless, possible systematic uncertainties may arise from selection effects that depend on the specific observables \citep{wu22} that could select systems in particular dynamical states and accretion rates \citep{shin23}. 

This paper is organised as follows. In Sec.~\ref{sec.data}, we present the photometric properties of the KiDS-DR3 data, and in Sec.~\ref{sec.measure}, we measure the stacked excess surface mass density signal in different redshift and amplitude bins. In Sec.~\ref{sec.model}, we introduce the density profiles used to model the cluster weak-lensing data, and in Sec.~\ref{sec.splashback}, we show our results on the characterization of the splashback radii of our cluster sample, together with the comparison with previous works. Finally, in Sec.~\ref{sec.summary}, we summarize and discuss our results. 

The cosmological model used in this work assumes a flat $\Lambda$CDM universe \citep{planck18}, in particular considering $\Omega_M=0.3$, $h=0.7$, and $\sigma_8=0.811$.

\section{Data set}
\label{sec.data}

We use the data of the Kilo-Degree Survey (hereafter KiDS) \citep{dejong13} collaboration, an optical wide-field imaging survey that has mapped the galactic sky in two stripes (an equatorial one, KiDS-N, and one centred around the South Galactic Pole, KiDS-S) and in four broad-band filters ({\em u, g, r, i}).  
The photometric data are taken with the 268 Megapixels OmegaCAM wide-field imager \citep[][]{omegacameso11}, composed by a mosaic of 32 science CCDs and located at the Very Large Telescope (VLT) Survey Telescope (VST). 
This is an ESO telescope of 2.6 meters in diameter located at the Paranal Observatory \citep[for further technical information about
VST see][]{capaccioli11}. 
The VST-OmegaCam field of view of $1\text{ deg}$ corresponds to about $12.5 \text{ Mpc}/h$ at the median redshift of the considered galaxy sample, $z\sim0.35$.
The principal scientific goal of KiDS is to exploit weak-lensing and photometric redshift measurements to map the large-scale matter distribution of the Universe and derive constraints on the main cosmological parameters.

In particular, here we use the KiDS-DR3 catalogue \citep{dejong17} and the AMICO cluster sample \citep{bellagamba18}.
This comprises about \num{100000} sources per square degree, resulting in about 50 million sources over the full survey area. 
KiDS-DR3 has a sky coverage of approximately $447 \deg^2$, composed of 440 survey tiles. It includes photometric redshifts and the corresponding probability distribution functions, a globally improved photometric calibration with respect to the previous releases, weak-lensing shear catalogues \citep{hildebrandt17}, and lensing-optimised image data. Source detection, positions, and shape parameters used for weak-lensing measurements are all derived from the $r$-band images, while magnitudes are measured in all filters using forced photometry.

The $440$ survey tiles of DR3 mostly cover a small number of large contiguous areas. This enables a refinement of the photometric calibration that exploits both the overlap between observations within a filter, as well as the stellar colours across filters \citep{dejong17}.
\footnote{The data products that constitute the main DR3 release (stacked images, weight and flag maps, and single-band source lists for 292 survey tiles, as well as a multi-band catalogue for the combined DR1, DR2, and DR3 survey area of 440 survey tiles), are released via the ESO Science Archive and they are also accessible via the Astro-WISE system (\href{http://kids.strw.leidenuniv.nl/DR1/access\_aw.php}{\rm http://kids.strw.leidenuniv.nl/DR1/access\_aw.php}) and the KiDS website \href{http://kids.strw.leidenuniv.nl/DR3}{\rm http://kids.strw.leidenuniv.nl/DR3}.}

The original properties of photometric redshifts (photo-$z$) of KiDS galaxies are described in \cite{kuijken15} and \cite{dejong17}. 
The photo-$z$s were extracted with BPZ \citep{benitez00, hildebrandt12}, a Bayesian photo-$z$ estimator based upon template fitting from the 4-band ({\em u, g, r, i}). Furthermore, BPZ returns a photo-$z$ posterior probability distribution function, which the adopted cluster finder (described in the next section) fully exploits. 
When compared with spectroscopic redshifts from the Galaxy And Mass Assembly \citep[GAMA,][]{liske15} spectroscopic survey of low redshift galaxies, the resultant accuracy is $\sigma_z\sim0.04(1+z)$, as shown in \cite{dejong17}.

\subsection{AMICO KiDS-DR3 cluster catalogue}

Galaxy cluster candidates are identified by the Adaptive Matched Identifier of Clustered Objects (\texttt{AMICO}) algorithm \citep{bellagamba18,maturi19} -- one of the two cluster finders selected by the Euclid Collaboration \citep{adam19} -- starting from the photometric galaxy catalogue and the so-called GaAP magnitudes, obtained by homogenising the PSF in the different bands.

The KiDS-DR3 galaxy catalogue provides the spatial coordinates, the two arcsec aperture photometry in $u$, $g$, $r$, $i$ bands and photometric redshifts for all systems down to the 5 $\sigma$ limiting magnitudes of 24.3, 25.1, 24.9, and 23.8 in the four bands, respectively.  
In the cluster identification process, it was avoided the explicit use of galaxy colours to minimise the dependence of the selection function on the presence (or absence) of the red sequence of cluster galaxies \citep{maturi19}. 

The cluster catalogue used in this work is the same one exploited in previous works \citep{bellagamba18,bellagamba19,radovich20,puddu21,lesci22b,ingoglia22,romanello24} and validated by \cite{maturi19}.  
All systems belonging to the KiDS-DR3 footprints that are heavily affected by satellite tracks, haloes created by bright stars, and images or artefacts are rejected \citep{maturi19}. The method used to assess the quality of the detections exploits realistic mock catalogues constructed from the real data themselves with the data-driven approach implemented in the Selection Function extrActor (SinFoniA), as described in Section 6.1 of \citet{maturi19}. These mock catalogues are also used to estimate the uncertainties on the quantities characterising the detections, as well as the purity and completeness of the entire sample. Moreover, we considered cluster detections with a signal-to-noise $S/N > 3.5$ within the redshift range $0.1\le z\le 0.6$, for a final sample of 6962 galaxy clusters. Objects at $z<0.1$ are discarded because of their low lensing power, while those at $z>0.6$ are excluded because the background galaxy density in KiDS-DR3 data is too small to allow a robust weak-lensing analysis. 
From the mock realizations, which account for the original galaxy data set, masks, photo-z uncertainties, and the clustering of galaxies \citep{maturi19}, we have estimated a purity of  90\% in the first amplitude bin and 100\% in the others; on the other hand, the completeness is approximately 75\% in the first bin and moves toward 100\% for higher amplitude values \citep{lesci22}.

For each cluster, we conservatively select the galaxy population, excluding the galaxies whose most likely redshift $z_{\rm s}$ is not significantly higher than the lens one $z_{\rm l}$ \citep{sereno17}: 
\begin{equation}
  z_{\rm s,min} > z_{\rm l} + \Delta z,
\end{equation}
where $z_{\rm s, min}$ is the lower bound of the region, including the 2 $\sigma$ of the probability density distribution $p(z)$ and $\Delta z$ is set to 0.05, similar to the typical uncertainty on photometric redshifts in the galaxy catalogue and well larger than the uncertainty on the cluster redshifts \citep{maturi19}.

The \texttt{AMICO} cluster search starts by convolving the 3D galaxy distribution with a redshift-dependent filter and creating a corresponding amplitude map, where every peak constitutes a possible detection. To each possible candidate, the algorithm returns angular position, redshift, $S/N$, and the signal amplitude $A$ -- which scales with the cluster richness -- defined as:
\begin{equation}
A(\theta_c,z_c)\equiv\alpha^{-1}(z_c)\sum_{i=1}^{N_{gal}}\frac{M_c(\theta_i-\theta_c,m_i)p_i(z_c)}{N(m_i,z_c)}-B(z_c), 
\label{AMICOAmpl}
\end{equation}
where $\theta_c$ are the cluster sky coordinates, $M_c$ is the cluster mass model (i.e. the expected density of galaxies per unit magnitude and solid angle) at the cluster redshift $z_c$, $N$ quantifies the noise distribution, $p_i(z)$, $\theta_i$, and $m_i$ indicate the photometric redshift distribution, the sky coordinates and the magnitude of the $i$-th galaxy, respectively; the parameters $\alpha$ and $B$ are redshift-dependent functions providing the normalisation and the background subtraction, respectively; whose values can be found in \citet{maturi19}.

A Schechter luminosity function \citep{schechter76} and a projected NFW \citep{navarro96} radial density profile describe the cluster model $M_c$. 
For each galaxy, labelled with  subscript $i$, the probabilistic membership association to the $j$-th detection is defined as:
\begin{equation}
P(i\in j)\equiv P_{f,i}
\frac{A_jM_{c,j}(\theta_i-\theta_j,m_i)p_i(z_j)}{A_jM_{c,j}(\theta_i-\theta_j,m_i)p_i(z_j)+N(m_i,z_j)}
, \label{Member prob.}
\end{equation}
where $A_j$, $\theta_j$ and $z_j$ are the amplitude, the sky coordinates and the redshift of the $j$-th detection, respectively. $P_{f,i}$ is the probability of the $i$-th galaxy belonging to the field before the $j$-th detection is defined, accounting for the possible multiple associations of each galaxy to more than one cluster. 
For the specific application to KiDS, the adopted filter includes galaxy coordinates, $r$-band magnitudes and the full photometric redshift distribution $p(z)$.

The distributions in redshift and amplitude of this cluster sample are shown in \cite{bellagamba19}. In the following analysis, we will divide the sample in three redshift bins:
\begin{itemize}
\item $0.1\le z<0.3$, with 2265 objects;
\item $0.3\le z<0.45$, with 2332 objects;
\item  $0.45\le z\le0.6$, with 2365 objects.
\end{itemize}
Those bins have been chosen to have almost the same number of clusters and numerous enough to study the amplitude trends in each of them.

\subsection{KiDS-DR3 galaxy sources for weak lensing analysis}

The original shear analysis of the KiDS-DR3 data has been described in \cite{kuijken15} and \cite{hildebrandt17}. 
The shape measurements have been performed with {\it lens}fit \citep{miller07,miller13} and successfully calibrated for the KiDS-DR3 data in \citet{fenechconti17}.

In this work, we use $r$-band data for shape measurements, selecting those with the best-seeing properties and the highest source density. 
The error on the multiplicative shear calibration, estimated from simulations with {\it lens}fit and benefiting from a self-calibration, is of the order of 1\% \citep{hildebrandt17}. 
The final KiDS-DR3 catalogue includes shear measurements for about $15$ millions of galaxies, with an effective number density of $n_{\rm eff}=8.53$ galaxies arcmin${^{-2}}$ \citep[as defined in][]{heymans12b}, over a total effective area of $360 \deg^2$.

\section{Measuring the excess surface mass density profile}
\label{sec.measure}

The surface mass density $\Delta\Sigma$ allows us to characterise the projected density profile of the observed clusters. It can be calculated knowing the lens and source redshifts, and the shear. However, we remind that the shear signal $\gamma$ is inaccessible from observed data. 
What is typically measured from the background galaxy ellipticities instead is an estimate of the reduced shear $g = \gamma/(1-\kappa)$, where $\kappa$ represents the lensing convergence \citep{bartelmann01}. In the weak-lensing regime where the convergence is much smaller than unity, we have $\gamma \simeq g$.

Intrinsic alignment (IA) of galaxies, when galaxy shapes are correlated with the underlying gravitational tidal field, can contaminate the lensing signal (II).  
Furthermore, background galaxies experience a shear caused by the foreground tidal gravitational field, and if the foreground galaxy has an intrinsic ellipticity that is linearly correlated with this field, shape, and shear are correlated (GI). 
Following the intrinsic alignment model \citep{bridle07,heymans12}, the power spectra of II and GI terms can be computed from the matter power spectrum, with a coefficient of proportionality that depends on the total matter density of the Universe and is inversely proportional to the linear growth factor. 
In cosmic shear analyses, depending on the number density of sources and their redshift distribution, IA is an important source of contamination and needs to be properly modelled in order to derive unbiased cosmological parameters \citep{asgari21,fishbacher23}. In the non-linear regime, that is the case of cluster lensing, until now, the search for intrinsic alignment signal has been relatively uncertain. 
The signal could be important for scales larger than 10 $h^{-1}$Mpc; however, \citet{chisari14} have shown that the IA signal for stacked clusters, as in our study, is consistent with zero. 

Calculating the tangential component of the shear signal $\gamma_t$ of background sources relative to the cluster centre, we can compute the excess surface mass density as:
\begin{equation}
\Delta \Sigma (r) = \bar{\Sigma}(<r) - \Sigma(r) \equiv \Sigma_{\rm
  crit} \gamma_t,
  \label{eqDeltaSigma}
\end{equation}
where $\Sigma(r)$ represents the surface mass density of the lens at distance $r$ and $\bar{\Sigma}(<r)$ its mean within $r$, written as:
\begin{equation}
 \bar{\Sigma}(<r) = \dfrac{2}{r^2} \int_0^r \mathrm{d}R\,R\,\Sigma(R)  .
  \label{eqEnclosedSigma}
\end{equation}
$\Sigma_{\rm crit}$ indicates the critical surface density \citep{bartelmann01}, expressed as:
\begin{equation}
  \Sigma_{\rm crit} = \dfrac{c^2}{4 \pi G} \dfrac{D_{\rm s}}{D_{\rm l} D_{l\rm s}}\,,
  \label{eq_scrit}
\end{equation}
with $D_{\rm l}$, $D_{\rm s}$, and $D_{\rm ls}$ being the
observer-lens, observer-source, and source-lens angular diameter distances, respectively, while $c$ represents the speed of light and $G$ is the Newton gravitational constant.

From the galaxy source catalogue, we then construct the excess surface mass density profile at a distance $r_j$ from the following relation:
\begin{equation}
  \Delta \Sigma (r_j) = \left( \dfrac{ \sum_{i \in j} \left( w_i \Sigma^{-2}_{\mathrm{crit},i} \right)  \gamma_{t,i}\Sigma_{\mathrm{crit}, i}}{\sum_{i \in j} \left( w_i \Sigma^{-2}_{\mathrm{crit},i}\right) } \right)  \dfrac{1}{1 + K(r_j) }\,,
\end{equation}
where $w_i$ indicates the weight assigned to the measurement of the source ellipticity and $K(r_j)$ the average correction due to the multiplicative noise bias in the shear estimate as in Eq.~(7) by \citet{bellagamba19}, which accounts for the shape noise. 
To compute the critical surface density for the $i$-th galaxy, we use the most probable source redshift as given by BPZ. However, it is worth underlining that due to the uncertainty of the photometric redshift and the anisotropy of $\Sigma_{{\rm crit}, i}$ with respect to the redshift,  multiplicative biases in the lensing profile may not be negligible \citep{mcclintock19}. 
In our KiDS-DR3 sample, we are confident that this quantity is quite small due to the excellent quality of the photometric redshift and negligible with respect to the statistical uncertainty. 

We stack clusters in amplitude and redshift bins, adopting the following criterion:
\begin{equation}
  \Delta \Sigma_N(r_j) = \dfrac{\sum_{n\in N} W_{n,j} \Delta \Sigma_{n}(r_j)
  }{\sum_{n \in N} W_{n,j}}\,,
\end{equation}
where $N$ represents the bin in which we perform the stacking and $W_{n,j}$ indicates the weight for the $j$-th radial bin of the $n$-th cluster:
\begin{equation}
 W_{n,j} = \sum_{i \in j} w_i \Sigma^2_{\mathrm{crit},i}\,.
\end{equation}

\begin{figure*}
\centering
\includegraphics[width=\hsize]{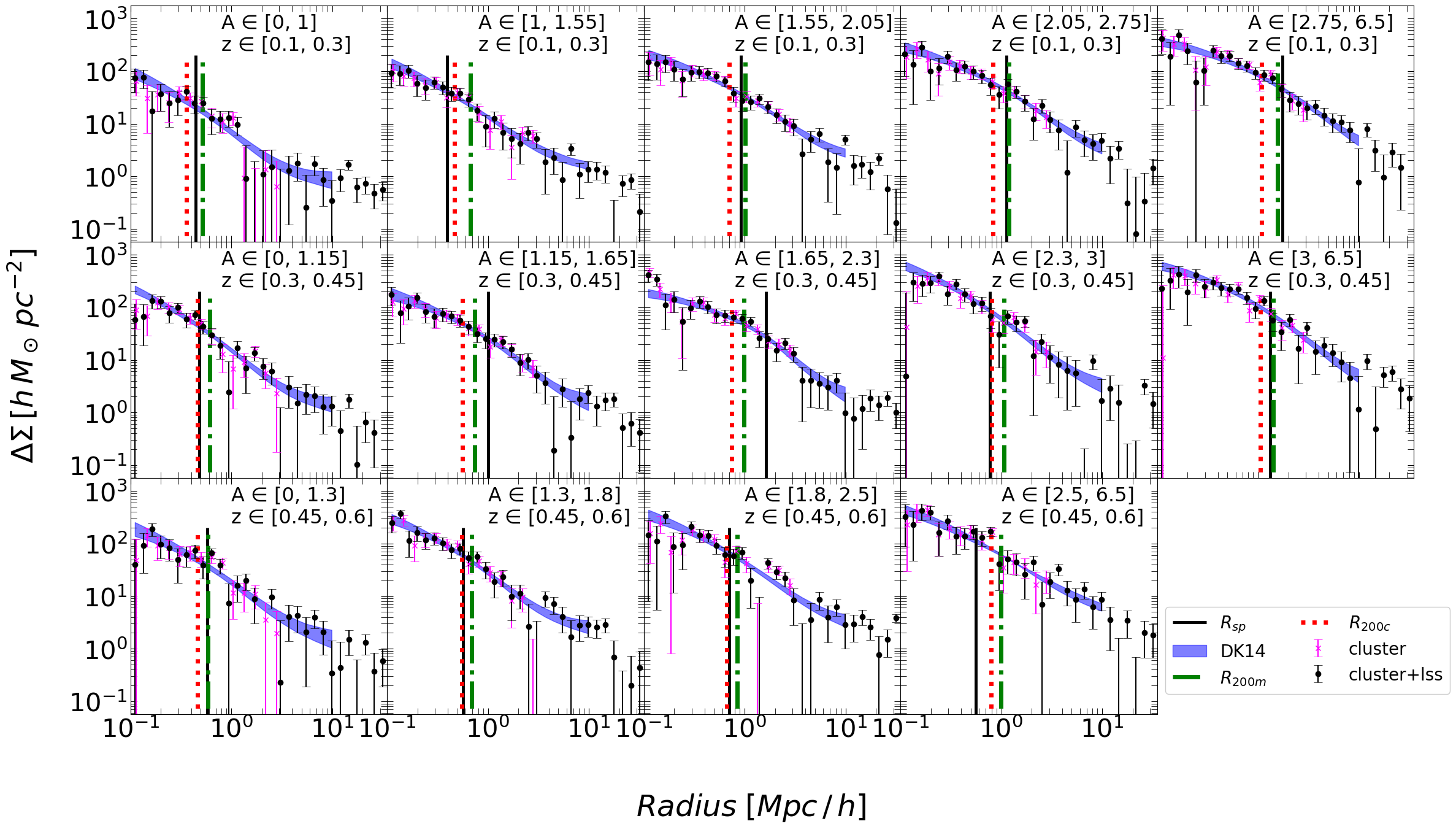}
\caption{Excess surface mass density data and models. 
The data points, with the corresponding error bars, represent the stacked measurements in different amplitude and redshift bins, as labelled in each panel.  
Magenta crosses and black-filled circles refer to the expected signals from the cluster region and from the cluster plus the large-scale structure (lss) contributions, respectively. 
In the latter case, we also subtract the signal around 10,000 random realisations of the cluster sample.
Blue-shaded regions enclose the 16th and 84th percentiles of the best-fit model, adopting DK14 \citep{diemer14}, to the black data points up to 12 $h^{-1}\mathrm{Mpc}$. 
The red-dotted, green-dashed and black solid vertical lines mark the location of different cluster radii: $R_{200c}$, $R_{200m}$m  and the splashback radius $R_{sp}$, respectively.} 
\label{figdeltaSmodels}
\end{figure*}

In Fig.\ref{figdeltaSmodels}, we display the stacked excess surface mass density in different redshift and amplitude bins measured from our data sets. 
In particular, the magenta crosses cover the cluster region and correspond to 14 radial bins from the \texttt{AMICO}-defined centres up to $R=3.16 Mpc/h$; the black-filled circles instead represent the cluster+lss (large-scale structure) part, which consists of 30 bins up to 35 Mpc/$h$. 
For this latter case, as done in \cite{giocoli21}, we remove the signal around random points within the survey masks, generating 10,000 random realizations of the total number of considered clusters in the corresponding redshift bins. 
This allows us to keep all the observational systematic uncertainties of the survey and the masks under control when modelling the lensing signal at large distances from the cluster centres. 
In addition, subtracting the signal around random points gives us the possibility to statistically remove contaminant large-scale structures that can bias the cluster lensing signal at large distances from the centre. The shaded blue region encloses the 18th and 84th percentiles of the best-fit model using the \citet{diemer14} that we will introduce in Sec.~\ref{secdk14}, as well as the characterization of the splashback radius displayed with the black solid vertical line.

In order to derive the best-fit model for the data, we use Bayesian inference analysis methods. 
From the data-vector $\vec{D}$, constructed stacking the excess surface mass density profiles of different clusters in a given amplitude and redshift bin, we can derive the posterior probability of a set of model parameters $\vec{\theta}$ using the Bayes' theorem:
\begin{equation}
 P(\vec{\theta}| \vec{D}) = \dfrac{\mathcal{L}(\vec{D}|\vec{\theta})P(\vec{\theta})}{E(\vec{D})}\,,
\end{equation}
where $\mathcal{L}(\vec{D}|\vec{\theta})$ represents the likelihood function of the data-vector given the model parameters, $P(\vec{\theta})$ the prior probability, and  $E(\vec{D})$ indicates the evidence of the data-vector.
For the likelihood we assume a Gaussian distribution:
\begin{equation}
\mathcal{L} \propto \exp \left( - \dfrac{1}{2} \chi^2 \right)\,,
\end{equation}
where the $\chi^2$ dependence on the data-vector and the model can be written as:
\begin{equation}
\chi^2 = \sum_{i,j}  \left[ \Delta \Sigma_{\rm data} -  \Sigma_{\rm model}\right]_i \,C_{ij}^{-1}\, \left[ \Delta \Sigma_{\rm data} -  \Sigma_{\rm model}\right]_j\,
\end{equation}
where $i$ and $j$ run on the radial bins, and $C_{ij}$ is the covariance matrix. 
 To compute the latter, we use 10,000 bootstrap realizations for each stacked cluster binned sample, following the approach of  \citet{giocoli21}. Our analysis reveals that the dominant terms are the diagonal ones, which are modulated by the shape noise variance in the logarithmically spaced bins, confirming the results obtained by \citet{bellagamba19}.
\section{Modelling the weak-lensing data}
\label{sec.model}

In this work, we adopt two models to characterise the clusters' profiles, using the measured excess surface mass density profiles of clusters in different amplitude and redshift bins presented in the previous Section. 
As done in \citet{sereno17,sereno18a,giocoli21,lesci22}, we first model the sample using the \citet[][hereafter BMO]{baltz09} function plus the cosmological large-scale term as the reference halo model. Then, we compare the mass derived using the BMO model to the results obtained with the \citet{diemer14} profile (hereafter DK14), which we will use to characterise the splashback radius. We now introduce both models in detail.

\subsection{BMO profile with cosmological large-scale contribution}

\label{sub:1halo}
In the BMO model \citep{baltz09}, the cluster main halo is modelled with a smoothly truncated Navarro-Frenk-White \citep[NFW,][]{navarro97} density profile:
\begin{equation}
  \rho(r)=\frac{\rho_{\rm s}}{\left(r/r_{\rm s}\right)\left(1+r/r_{\rm s}\right)^2}\left(\frac{r_{\rm t}^2}{r^2+r_{\rm t}^2}\right)^2, \label{trunc.NFW}
\end{equation}
where $\rho_{\rm s}$ represents the typical matter density within the scale radius $r_{\rm s}$, and $r_{\rm t}$ indicates the truncation radius that we express in terms of the halo radius $R_{\rm 200m}$, that is the radius enclosing $200$ times the mean background density of the Universe at the considered redshift, namely $\rho_{\rm m}(z)$. Specifically, in Eq.\ \eqref{trunc.NFW}, $r_{\rm t} \equiv t\,R_{\rm 200m}$, where $t$ is defined as the truncation factor. The scale radius is parameterized as $r_{\rm s}\equiv R_{\rm 200m}/c_{\rm 200m}$, where the concentration $c_{\rm 200m}$ is correlated with halo mass and redshift, depending on the halo mass accretion history \citep{maccio07,neto07,maccio08,zhao09,giocoli12b}.  
The total mass enclosed within the radius $R_{\rm 200m}$, referred to as $M_{\rm 200m}$, can be seen as the normalisation of the model and a mass proxy of the true enclosed mass of the dark matter halo hosting the cluster \citep{giocoli12a}.
This truncated version of the NFW model has been deeply tested in simulations \citep{oguri11b,giocoli24}, which demonstrated that it describes the cluster profiles more accurately than the original NFW profile, removing the non-physical divergence of
the total mass at large radii.  

Moreover, the BMO model accurately describes the transition between the cluster's main halo and the 2-halo contribution \citep{cacciato09,giocoli10b,cacciato12}, providing less biased estimates of mass and concentration from shear profiles \citep{sereno17}.  By neglecting the truncation, the mass would be underestimated, and the concentration overestimated \citep{giocoli24}.  

As we are considering only stacked shear profiles, the spherical symmetric model provides a reliable description, given that the intrinsic halo triaxialities are statistically averaged out.
We characterise the truncation scale as in \citet{bellagamba19} and following the results by \citet{oguri11b}: considering $M_{\rm 200c}$ they defined the truncation radius as $r_t=3\times R_{\rm 200c}$. 
We convert this truncation radius in terms of $R_{\rm 200m}$ using \texttt{colossus} \citep{diemer18}, assuming a \citet{diemer19} mass-concentration relation,  and thus deriving the ratio between $R_{\rm 200m}$ and $R_{\rm 200c}$. 

At scales larger than $R_{\rm 200m}$, the shear signal originates from the correlated matter distribution around the galaxy clusters.  In practice, the 2-halo term characterises the cumulative effects of the large-scale structures in which galaxy clusters are embedded. The uncorrelated matter distribution along the line-of-sight produces only a modest random contribution to the stacked shear signal, accounted for in the error budget.  
We model the 2-halo term following the recipe by \citet{oguri11}.
  
The total model for the excess surface mass density profile is the sum of the halo profile described by Eq. \eqref{sigma1halocen.+off} and the contribution due to the matter in correlated haloes that we can write as
\citep{oguri11,oguri11b,sereno17}:
  \begin{equation}
  \Delta\Sigma_{\rm 2h}(\theta;M_{\rm 200m},z) \!= \!\int
  \text{d}l\frac{l}{2\pi}J_2(l\theta)
  \frac{\bar{\rho}_{\rm m}(z)b_{\rm h}(M_{\rm 200m};z)}{(1+z)^3D_l^2(z)}P_{\rm lin,m}(k_{\rm l};z),
  \label{surf.dens.2haloterm}
  \end{equation}
where $z$ represents the cluster redshift, estimated using photometric data, as provided by \texttt{AMICO}. 
The other terms in Eq.~(\ref{surf.dens.2haloterm}) are summarised as follows:
\begin{itemize}
  \item $\theta$ is an angular scale;
  \item $J_2$ is the Bessel function of the second type. This is a function of $l\theta$, where $l$ is the integration variable and the momentum of the wave vector $k_{\rm l}$ for the linear power spectrum of matter fluctuations $P_{\rm lin,m}$;
  \item $k_{\rm l}=l/((1+z)D_{\rm l}(z))$ indicates the wave vector module;
  \item $\bar{\rho}_{\rm m}(z)$ represents the mean cosmic background density at the lens redshift;
  \item $P_{\rm lin,m}(k_{\rm l};z)$ is the linear matter power spectrum computed according to the \citet{eisenstein99} transfer function, which is accurate enough given our current measurement uncertainties. We also tested an alternative 
  prescription using the CAMB model \citep{camb}, finding negligible differences;
  \item $b_{\rm h}(M;z)$ is the halo bias, for which we adopt the \citet{tinker10} model, which has been successfully tested and calibrated using a large data set of numerical simulations. 
  \end{itemize} 

In what follows, we will generally term the full 1h+2h term as the BMO model: the analytical form on small scales is described by the profile in Eq.~\ref{trunc.NFW} and on large scale by Eq.~\ref{surf.dens.2haloterm}.

\subsection{DK14 profile}
\label{secdk14}
In the DK14 model \citep{diemer14}, the total matter
density present in collapsed structures is constructed using the Einasto profile \citep{einasto65}, which has an extra parameter compared to NFW that captures the slope variation in the inner region. DK14 used the Einasto model to describe the \emph{orbiting material} in the internal part of the halo and included a new \emph{infall} term to characterize the matter at larger distances from the centre. 
The DK14 profile can then be read as:
\begin{align}
    \label{dk14}
    &\rho(r) = \rho_{\rm s} \, \exp\bigg\{-\frac{2}{\alpha}\,\bigg[\bigg(\frac{r}{r_{\rm s}}\bigg)^\alpha -1\bigg]\bigg\} \, \,f_{\rm trans} +\rho_{\rm outer},\;\mathrm{where:} \\
    & f_{\rm trans} = \bigg[1 + \bigg(\frac{r}{r_{\rm t}}\bigg)^\beta\bigg]^{-{\frac{\gamma}{\beta}}},  \label{trans}\\
    &\rho_{\rm outer} = \rho_{\rm m} \, \bigg[\, b_{\rm e} \,\bigg(\dfrac{r}{5R_{\rm 200m}}\bigg)^{-s_{\rm e}}+1 \,\bigg]. 
    \label{infal}
\end{align}
The first equation includes the Einasto profile that is multiplied by a transition or truncation function $f_{\rm trans}$ (see Eq.~\ref{trans}) that models the transition between the internal halo component and the infalling large-scale term $\rho_{\rm outer}$ (see Eq.~\ref{infal}). 
Like in the BMO profile, the transition term depends on the truncation radius $r_{\rm t}$ and on two other parameters: $\beta$, which describes how fast the slope changes around the truncation radius, and the slope $\gamma$. Eq.~\ref{infal}, namely the outer term, models the matter density distribution outside the halo radius $R_{200m}$, described by a power-law term that multiplies the mean background density of the Universe, $\rho_{\rm m}$. 
Following the formalism described in DK14, we define the pivot radius as $5$ times $R_{\rm 200m}$. 
The parameters $b_{\rm e}$ and $s_{\rm e}$ describe the normalisation and the slope of the power law, respectively; in particular, for $s_{\rm e} \gg 1$ the outer term reaches the background density at large distances from the centre. 

The model and the parameters, as described in DK14, have been implemented in the \texttt{CosmoBolognaLib}\footnote{\href{https://gitlab.com/federicomarulli/CosmoBolognaLib}{\rm https://gitlab.com/federicomarulli/CosmoBolognaLib}} \citep{marulli16} libraries used in this work.  We have set $\beta=4$, $\gamma = 4 \nu_{\rm vir}$, and $\alpha=0.155 + 0.0095 \nu_{\rm vir}^2$ \citep{gao08} (the inner slope of the Einasto profile) where $\nu_{\rm vir}=\sigma^2(M_{\rm vir})/\delta^2_{\rm c}(z)$ represents the peak height of the considered halo mass,  $\delta_{\rm c}(z)$ the critical linear theory overdensity required for spherical collapse divided by the growth factor and $\sigma^2(M_{\rm vir})$ the mass variance computed as integral of the linear matter power spectrum convolved with a top-hat window function with a scale equal to $R_{\rm vir}$. To be consistent with the definition by \citet{gao08}, we convert $M_{\rm 200m}$ to $M_{\rm vir}$ assuming an NFW profile with the considered mass and concentration; for the virial overdensity definition, we adopt the fitting function by \cite{bryan98}.

\subsection{The mis-centring: \emph{to be or not to be}}

When analysing the weak-lensing signal produced by galaxy clusters, an important source of uncertainty is the inaccurate identification of the lens centre. In this study, we assume the centres determined by \texttt{AMICO} in the detection procedure and do not account for mis-centring. This is motivated by the fact that the splashback radius is located at a large distance from the cluster centre and, therefore, should not be significantly affected by the density profile in the inner region.  
However, since the mass is a normalisation factor in the lensing models, we could expect an impact of mis-centring on the concentration estimate. 

In any case, our prescription can also account for the inaccurate centring by considering a second additional component in both BMO and DK14 models and following the formalism as described in \citet{viola15,johnston07,giocoli21}.
Our final model for the 1-halo term (either BMO or Einasto) can be written as the sum of a centred and an off-centred population:
 \begin{equation}
   \Sigma_{1h}(R)=(1-f_{\rm off})\Sigma_{\rm cen}(R)+f_{\rm off}\Sigma_{\rm off}(R).
   \label{sigma1halocen.+off}
 \end{equation} 
Therefore, our analysis of the 1-halo radial range uses a model that depends on four parameters: $M_{\rm 200m},c_{\rm 200m},\sigma_{\rm off}$, and $f_{\rm off}$; and when fixing $\sigma_{\rm off}$ (the scale length) and $f_{\rm off}$ (fraction of off-centred systems) to zero it depends only of the halo mass and concentration.\footnote{It is worth noticing that in the 1-halo term modelling, using the BMO profile, \citet{giocoli21} have found $\langle f_{\rm off}\rangle =$ $0.2\pm 0.14$ and $\langle \sigma_{\rm off}\, [h^{-1}\mathrm{Mpc}]\rangle =0.25 \pm 0.13 $, the last one dominated by the smallest radial range considered.}
\begin{figure}
\centering
\includegraphics[width=\hsize]{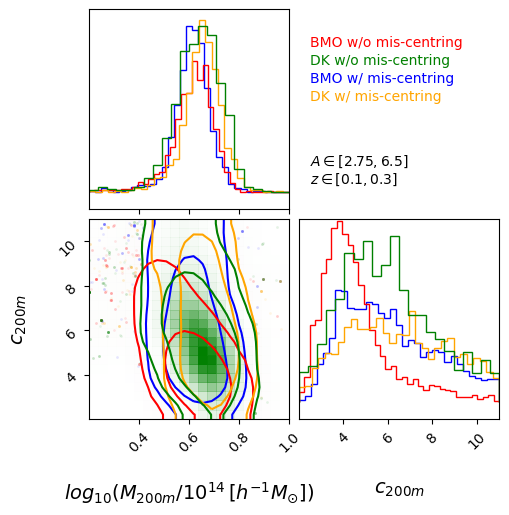}
\caption{Posteriors distribution in the recovered mass $\log_{10}\left(M_{\rm 200m}\right)$ and concentration $c_{\rm 200m}$ when modelling the cluster data, the magenta crosses in Fig.\ref{figdeltaSmodels} for the top right case corresponding to the first redshift bin and to the last amplitude bin. Red (green) and blue (orange) distributions show the posteriors of the recovered mass and concentration using the BMO (DK14) model without and with the mis-centring uncertainty distributions, respectively.
\label{figposteriors}}
\end{figure}

In our study, we neglect the measurements below radial separations of $0.2$ Mpc$/h$ mainly for three reasons.  Firstly, from an observational point of view, the uncertainties in the measure of photo-$z$s and shear close to the cluster centre are large because of the contamination due to the higher density of galaxies in which blending deteriorates the shape and photometric measures. 
Secondly, the shear signal analysis in close proximity to the cluster centre is sensitive to the BCG contribution to the matter distribution and to deviations from the weak-lensing approximation used in the profile model. 
Lastly, this choice mitigates the miscentring effects. Thus, neglecting small-scale measurements minimises the systematics uncertainties, possibly affecting the estimation of the concentration $c_{\rm 200m}$, which would otherwise be overestimated - being degenerate with the $\sigma_{\rm off}$ and $f_{\rm off}$ parameters.
In Tab.\ref{tabpars_clusters}, we report the free parameters of our models and the priors we adopt in our MCMC analyses when modelling our observational data of the cluster region as described by the magenta crosses in Fig.\ref{figdeltaSmodels}.

We want to underline that since the goal of this work is to characterise the splashback radius, expected to be located a very large distance from the centre and close to $R_{\rm 200m}$, as our reference results, we consider the model without mis-centering.
\begin{table}
\centering	
\caption{Adopted prior parameters for the BMO and DK14 models when modelling the cluster data with 0.2 $h^{-1}$Mpc $< r <$ 3.16 $h^{-1}$Mpc from the centre.}
\label{tabpars_clusters}
\begin{tabular}{lc|r|r}
\hline
$ $ & $  $ & BMO & DK14 \\
\hline
$\log_{10} \left( M_{\rm 200m}  h / M_{\odot}\right)$ & $[12.5, 15.5]$ & \checkmark & \checkmark \\
$c_{\rm 200m}$ & $[1, 20]$ & \checkmark & \checkmark \\
when free: $f_{\rm off}$ & [0,0.5] & \checkmark & \checkmark \\ 
when free: $\sigma_{\rm off} [h^{-1}$Mpc$]$ & [0,0.5] & \checkmark & \checkmark \\ 
$r_{{\rm t}}/R_{\rm 200m}$ & $[0.8,5]$ & -- & \checkmark \\ 
$b_{\rm e}$ & $[0.1,4]$ & -- & \checkmark \\ 
$s_{\rm e}$ & $[0.5,2]$ & -- & \checkmark \\ 
 \hline 
\end{tabular}
\end{table}
To support our choice, we show an example in Fig.\ref{figposteriors}, where we present the posterior distributions on the recovered mass $\log_{10}M_{\rm 200m}$ and concentration $c_{\rm 200m}$ using our two models when switching on and off the mis-centring terms.
The red and green and the blue and orange distributions refer to the BMO and DK14 without and with the mis-centring terms, respectively. 
The case displayed, representative of all redshift and amplitude bins, corresponds to the top right panel of Fig.\ref{figdeltaSmodels}: namely $z\in[0.1,0.3]$ and $A\in[2.75,6.5]$, for the cluster case data and referring to the magenta crosses.
From Fig.\ref{figposteriors}, we notice that the recovered masses in the four cases are all consistent well within the 1 $\sigma$ confidence region. 
Also, the concentrations are fairly consistent but there is an evident degeneracy with the mass. 
While the models without the mis-centring term predict a smaller concentration parameter and have narrower posterior distributions, the cases which include the mis-centring terms tend to predict slightly larger concentrations -- consistent with the fact of having a larger number of free parameters.  
It is also worth noticing that since the DK14 model has more parameters left free than the BMO one, the posteriors in the concentration parameter are wider. 
In Tab.\ref{tabpars_chi2} we report, for each redshift and amplitude bin, the $\chi^2$ (and reduced $\chi^2$) for the BMO and DK14 model, accounting or not for the mis-centring terms. Both BMO and DK14 profiles well describe the cluster excess surface mass density (magenta crosses in Fig. \ref{figdeltaSmodels}); nevertheless, the DK14, having a more flexible inner slope, tends to have, on average, better goodness of fit. 

\begin{table*}
\centering	
\caption{$\chi^2$ for the different considered models when fitting the data of the cluster region up to 3.61 Mpc/$h$, where we have 14 data points. The corresponding $\chi^2/$ n. dof are also reported between brackets.}
\label{tabpars_chi2}
\begin{tabular}{cccccccc}
\hline 
redshift range & Amplitude range & BMO & DK14 & BMO  & DK14  \\ 
$z \in$ & $A\in$ &  &  & w/ miscent. &  w/miscent. & $N_{\rm cl}$ \\ 
\hline 
$[0.1,0.3]$ & $[0,1]$  & 12.049 (1.004) &  11.949 (1.328) &  11.614 (1.161)  &  11.498 (1.643) & 1066 \\
$[0.1,0.3]$ & $[1,1.55]$  & 8.740 (0.728)  &  6.930 (0.770) &  8.380 (0.838) &  7.291 (1.042) & 822 \\
$[0.1,0.3]$ & $[1.55,2.05]$  & 6.519 (0.543)  &  7.097 (0.788) &  6.805 (0.680) &  6.791 (0.970) & 240 \\
$[0.1,0.3]$ & $[2.05,2.75]$  & 7.834 (0.653)  &  7.590 (0.843) &  7.968 (0.797) &  8.090 (1.156) & 96 \\
$[0.1,0.3]$ & $[2.75,6.5]$  &  9.481 (0.790)  &  12.580 (1.398) & 9.097 (0.910) &  10.764 (1.538) & 41 \\

$[0.3,0.45]$ & $[0,1.15]$  &    16.806 (1.400) &  16.682  (1.854) & 16.900 (1.690)  & 17.540  (2.506) & 1090 \\
$[0.3,0.45]$ & $[1.15,1.65]$  & 2.973 (0.248) &   3.694 (0.410) &  3.691 (0.369) &  2.909 (0.416) & 762 \\
$[0.3,0.45]$ & $[1.65,2.3]$  &  8.496 (0.708) &   10.761 (1.196) &  8.351 (0.835) &  10.774 (1.540) & 339 \\
$[0.3,0.45]$ & $[2.3,3]$  &     13.125 (1.094) &  12.460 (1.384) &  13.013 (1.301) &  13.080 (1.869)& 98 \\
$[0.3,0.45]$ & $[3,6.5]$  &     11.936 (0.995) &  12.551 (1.395) &  11.780 (1.178) &  12.377 (1.768) & 43 \\

$[0.45,0.6]$ & $[0,1.3]$  &     21.372 (1.781) &  21.552 (2.389) & 20.470 (2.047)  &  20.799 (2.970) & 984 \\
$[0.45,0.6]$ & $[1.3,1.8]$  &   8.307  (0.692) &  7.821 (0.869 &  7.528 (0.753) &  6.789 (0.970) & 889 \\
$[0.45,0.6]$ & $[1.8,2.5]$  &   23.725 (1.977) &  22.525 (2.503) &  24.430 (2.443) &  23.010 (3.287)& 373 \\
$[0.45,0.6]$ & $[2.5,6.5]$  &   13.048 (1.087) &  13.854 (1.539) &  12.429 (1.243) &  13.287 (1.898) & 119 \\

 \hline 
\end{tabular}
\end{table*}

\subsection{Scaling relations}

Calibrating the scaling relations of galaxy clusters is an essential step in observational cosmology, that is in the context of using galaxy clusters as cosmological probes. 
Those relations describe the proportionality between observable quantities of galaxy clusters and their underlying physical properties. 
Typically, scaling relations depend on redshift, in particular on the dynamic state of galaxy clusters and the changing observational wavelength as we look back in time across the history of the Universe.
In this respect, finding the scaling relation that mildly evolves with redshift could represent a successful approach. 

\begin{figure}
\centering
\includegraphics[width=\hsize]{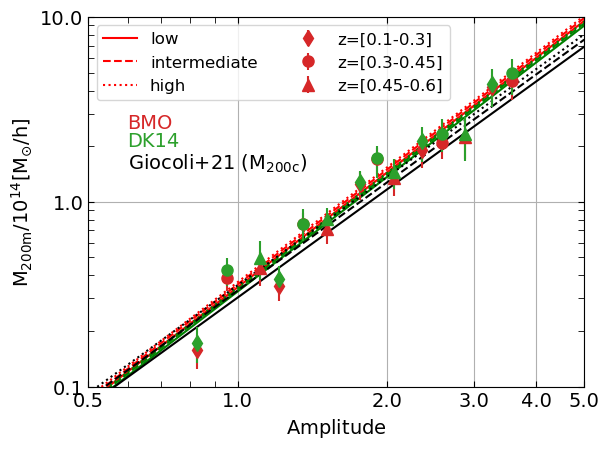}
\caption{Mass-amplitude relation calibrated using BMO and DK14 without accounting for the mis-centring uncertainties. The data points and the corresponding error bars exhibit the median and the 16th and 84th percentiles of the posterior $\log_{10}\left(M_{\rm 200m}\right)$ distributions. 
The corresponding coloured lines show the best-fit scaling relations in the three redshift bins. 
For comparison purposes, the black lines show the results by \citet{giocoli21} adopting a different overdensity mass definition $M_{200c}$.
\label{ScalingRel}}
\end{figure}

\begin{figure}
\centering
\includegraphics[width=\hsize]{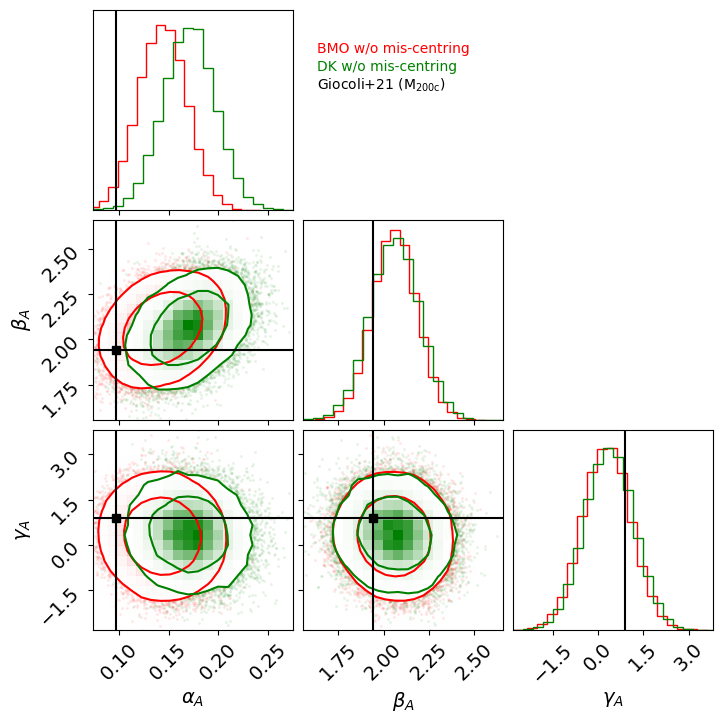}
\caption{Posterior distributions of the parameters of the mass-amplitude relation as in Eq.\ref{eq_mobs}. 
The contours show the 1 and 2 $\sigma$ confidence regions of the distributions for the BMO (red) and DK14 (green), both without accounting for the mis-centring of clusters. 
In Tab.\ref{tab_emceelinear}, we report the median values with the corresponding  16th and 85th percentiles for all cases. The black points indicate the median of the posteriors derived by \citet{giocoli21} when considering  the mis-centring terms and the BMO model in deriving $M_{200c}$.\label{ScalingRelPost}}
\end{figure}

In Fig.\ref{ScalingRel}, we show the relation between the \texttt{AMICO} amplitude $A$ and the weak-lensing recovered mass for all bins and considering the data from the cluster regions ($R<3.16\,\mathrm{Mpc}/h$). 
Red and green data points (coloured as in Fig.\ref{figposteriors}) show the results when adopting BMO and DK14 models without mis-centring, respectively. 
As discussed by \citet{sereno15c} and \citet{bellagamba19}, we model the trend using a linear relation between the logarithm of the two quantities:
\begin{equation}
\log \dfrac{M_{200m}}{10^{14} M_{\odot}/h} = \alpha_{\rm A} + \beta_{\rm A} \log
\dfrac{A}{A_{\rm piv}} + \gamma_{\rm A} \log \dfrac{E(z)}{E(z_{\rm piv})}\,,
\label{eq_mobs}
\end{equation}
where we set $A_{\rm piv}=2$, $z_{\rm piv}=0.35$ as in
\citet{bellagamba19} and $E(z) = H(z)/H_0$. 
For comparison, we show in black the results by \cite{giocoli21} when recovering $M_{200c}$.
In Fig.\ref{ScalingRelPost}, we show the posterior distributions of the scaling relation parameters displaying the 1 $\sigma$ and 2 $\sigma$ credibility regions.
We follow the same fitting procedure adopted in \citet{bellagamba18,giocoli21}, where the amplitude is computed through a lensing weighted average. Moreover, we account for the systematic errors in the covariance by summing the uncertainties for background selection, photo-$z$s, and shear measurements in quadrature. 
We assume uniform priors in the range $-1 <\alpha_{\rm A} < 1$, $0.1 <\beta_{\rm A} < 5$, and $-5 < \gamma_{\rm A} < 5$. From the figure, we can notice that there is a small positive degeneracy between the slope and normalization parameters.

In Tab.\ref{tab_emceelinear}, we summarise our results on the scaling relation parameters, as well as the cases where we include the mis-centring terms. 
From the table, it is worth underlining that the recovered parameters, with and without mis-centring for each given model are all consistent within the 16th and 84th percentiles. 
We can also notice the consistency between the scaling relation parameters adopting different models; this confirms that the concentration parameter degenerates with the mis-centring terms while $M_{\rm 200m}$ masses are all consistently recovered. 

From the figures, we notice that adopting as a mass proxy the weak-lensing derived mass enclosing 200 times the background density decreases the redshift dependence of the relation, considering also the uncertainties, in fact, we obtain $\lvert \gamma_{\rm A} \rvert$ much smaller than unity.

\begin{table}
\centering	
\caption{Scaling relation parameters of the mass-amplitude   relation (Eq.\ref{eq_mobs}). 
The values reported represent the median and 16th-84nd percentiles of the posterior distributions when modelling the cluster signal data using different models.}
\label{tab_emceelinear}
\begin{tabular}{lccc}
\hline
case & $\alpha_{\rm A}$ & $\beta_{\rm A}$ & $\gamma_{\rm A}$ \\ \hline
BMO w/o miscent. &   $0.144_{-0.024}^{+0.024}$ & $2.059_{-0.122}^{+0.123}$  & $ 0.291_{-0.794}^{+0.791}$ \\
DK14 w/o miscent. & $ 0.171_{-0.026}^{+0.026}$ & $2.062_{-0.136}^{+0.134}$ & $ 0.370_{-0.808}^{+0.805}$ \\
BMO w/ miscent. & $0.168_{-0.025}^{+0.024}$  & $1.975_{-0.131}^{+0.130}$ & $ 0.085_{-0.838}^{+0.819}$ \\
DK14 w/ miscent. & $0.197_{-0.025}^{+0.025}$  & $2.005_{-0.131}^{+0.132}$ & $ 0.363_{-0.820}^{+0.816}$ \\
\hline
\end{tabular}
\end{table}

\section{The splashback radius}
\label{sec.splashback}

\begin{figure}
\centering
\includegraphics[width=\hsize]{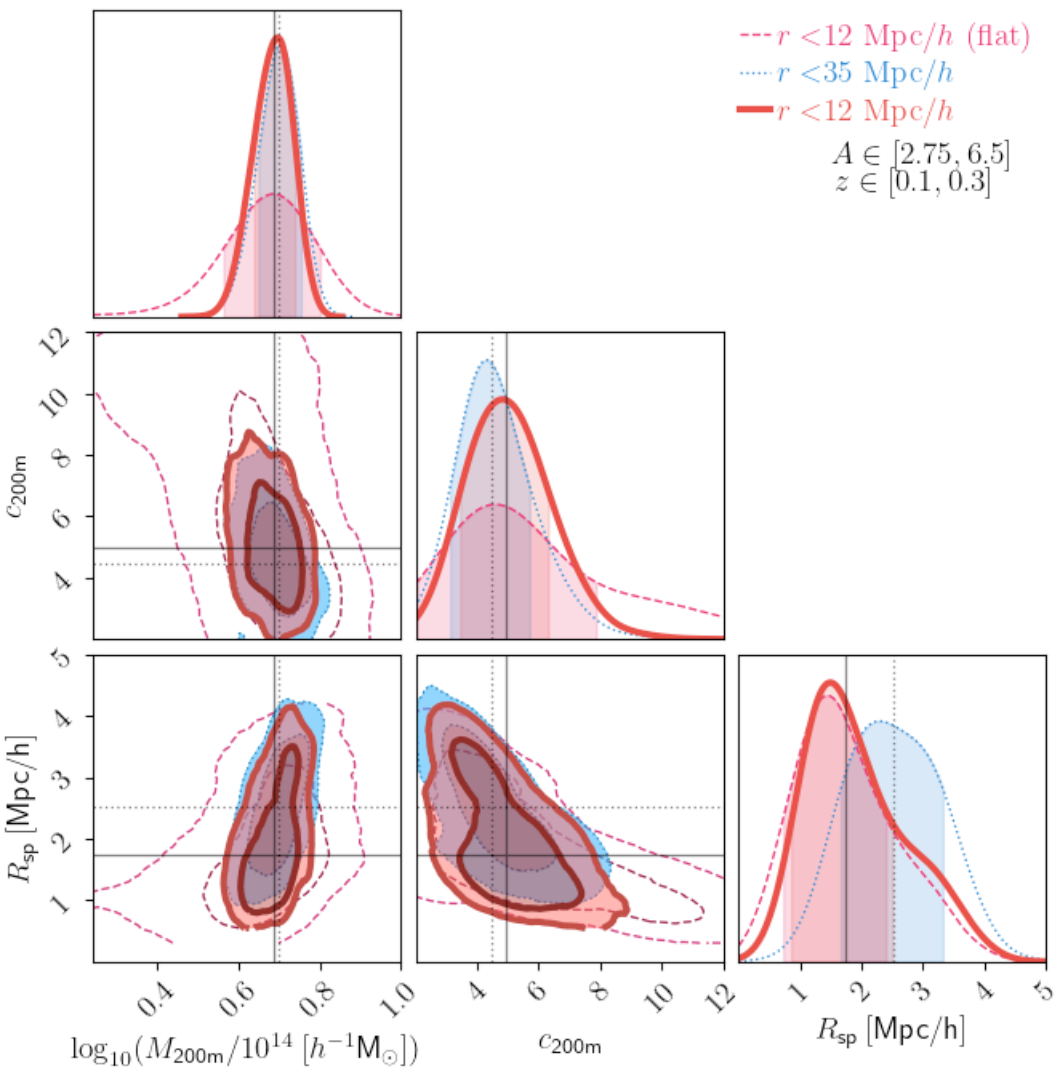}
\caption{\label{figDKposteriorsRsp} Posteriors distributions of the recovered mass concentration $c_{200m}$ and splashback radius $R_{sp}$, modelling the black filled circles as in Fig.\ref{figdeltaSmodels}, corresponding the same stacked cluster sample as in Fig.\ref{figposteriors}.
The red solid distribution shows our reference result that is modelling the data up to 12 Mpc/$h$ and adopting Gaussian priors for both the logarithm of the mass and the concentration. The blue-dotted distributions have been obtained modelling the data up to 35 Mpc/$h$, while the magenta dashed refers to the case in which we assume flat priors for $\log_{10}\left(M_{200m}\right)$ and $c_{200m}.$}
\end{figure}

In this section, we describe the characterisation of the splashback radius $R_{sp}$, from the \texttt{AMICO} KiDS-DR3 data,  obtained by modelling the excess surface mass density profile of the cluster+lss - see the black-filled circles in Fig.\ref{figdeltaSmodels}. 

We measure $R_{sp}$ from the profiles calculated in the previous Section using the DK14 model; $R_{sp}$ is calculated as the distance from the cluster centre where the logarithmic derivative of the density profile, $\mathrm{d}\log \rho(r)/\mathrm{d}\log r$, has its minimum value. 
This scale distinguishes two regions: the internal part, where galaxies are orbiting and bound to the cluster, and the external one, where material is infalling towards the cluster region along the cosmic web.
For the data vector, we consider two different radial ranges: $(i)$ the full data ($R\leq 35\,\mbox{Mpc}/h$), as also used by \cite{giocoli21,ingoglia22} to constrain cosmological parameters such as $\Omega_m$ and $\sigma_8$, and $(ii)$ data points only within $R\leq12\,\mbox{Mpc}/h$. 
The latter represents the scale up to which the infall term of the DK14 model has been built and tested in cosmological simulations. 
For the model parameter priors, we refer to Tab.\ref{tabpars_clusters}; in our fiducial result, we use Gaussian priors for the mass $\log_{10}(M_{\rm 200m})$ and the concentration $c_{\rm 200m}$ corresponding to the results as derived by modelling the cluster data, as discussed in the previous section. In Tab. \ref{tabpars_clusterslss}, we display the adapted priors for our reference case runs. 

As an example, in Fig.\ref{figDKposteriorsRsp}, we show the posterior distributions of the mass, concentration, and splashback radius for the systems in the top-right panel of Fig.\ref{figdeltaSmodels}, the same also shown in Fig.\ref{figposteriors}. 
The solid red (dotted blue) distributions show the 1 and 2 $\sigma$ confidence regions obtained by modelling the cluster+lss region up to 12 Mpc/$h$ (35 Mpc/$h$); the solid (dotted) black lines display the median values of the corresponding distributions. 
We can notice that when using data up to 35 Mpc/$h$, the splashback radius tends to be, on average, larger than when considering data up to 12 Mpc/$h$, for this stacked cluster sample. To show the impact of the priors, the dashed magenta distributions display the results adopting flat uniform priors for both the logarithm of the mass and the concentration; it is worth underlining that while the posterior distributions of the recovered mass and concentration are wider, the posterior distribution of the splashback radius is very similar to the case in which we assume Gaussian priors. 
The 16th and 84th percentiles of the best-fit model are displayed in shaded blue in all panels in Fig.\ref{figdeltaSmodels} for all amplitude and redshift bins. To compare to the DK14 model, we now continue to analyse our results calculated within 12 Mpc/h.

\begin{figure*}
\centering
\includegraphics[width=\hsize]{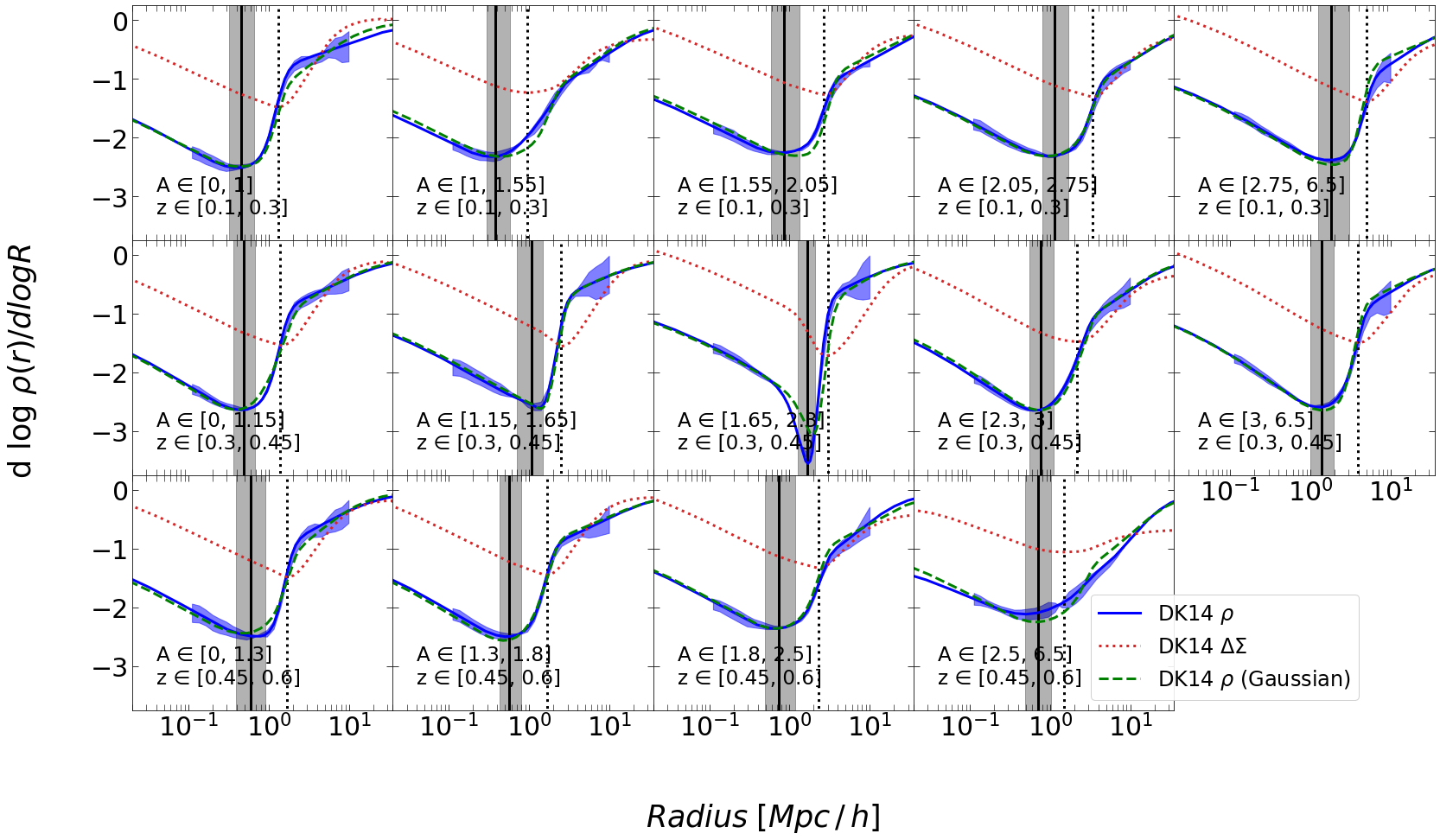}
\caption{Logarithmic density derivative of the best-fit model for different increasing redshift (from top to bottom)
and Amplitude (left to right) bins, as in Fig.\ref{figdeltaSmodels}. The solid blue and the dotted red curves show the 
logarithmic derivative of the three-dimensional density and projected excess surface mass density profiles computed 
using the median values of the posterior recovered distributions. The black vertical lines and the dashed grey regions 
indicate the location of the splashback radius -- corresponding to the minimum of the logarithmic 
derivative of the density profile -- with the 16th and 84th percentiles of the distributions. The vertical dotted lines 
show the position of the minimum of the logarithmic derivative of the $\Delta \Sigma$ profile.\label{figDlogrho} The dashed green curves, almost overlapping the blue ones,  show the results when assuming Gaussian priors on the parameters $\alpha_0$, $\alpha_{\nu}$, $\beta$ and $\gamma_0$ that
describe the Einasto profile and the transition term in the DK14 model.}
\end{figure*}

In Fig.\ref{figDlogrho}, we show the logarithmic derivative of the best-fit models -- constructed from the median values of the posterior distributions -- for each redshift and amplitude bin, ordered as in Fig.\ref{figdeltaSmodels}. The shaded regions display the propagation of the relative 1 $\sigma$ confidence region as computed from modelling the excess surface mass density in the radial range between 0.1 and 12 $\mbox{Mpc}/h$. 
The solid blue curves show the logarithmic slope variation of the three-dimensional density profile as a function of radius, computed from the median value of the posterior distributions using the \cite{diemer14} model. 
Black vertical lines indicate the splashback radius derived from the fit, with the corresponding 16th and 84th confidence intervals derived from the posterior distributions. 
The dotted red curves and black dotted vertical lines show instead the logarithmic derivative of the excess surface mass density profile  $\Delta \Sigma$ and the corresponding location of the minima.
From the figure, we can notice this is, on average, a factor of approximately 2.8 larger than the splashback radius computed from the three-dimensional density because of projection effects. \footnote{We have tested this also using \texttt{COLOSSUS} and constructing a composite profile with Einasto plus background and power-law for the external part (following the tutorial at this url \href{https://bdiemer.bitbucket.io/colossus/_static/tutorial_halo_profile.html}{\rm https://bdiemer.bitbucket.io/colossus/\_static/tutorial\_halo\_profile.html}), using the median values from our posterior distributions, and we found full agreement with our results, obtained from the \texttt{CosmoBolognaLib}.}

\begin{figure}
\centering
\includegraphics[width=\hsize]{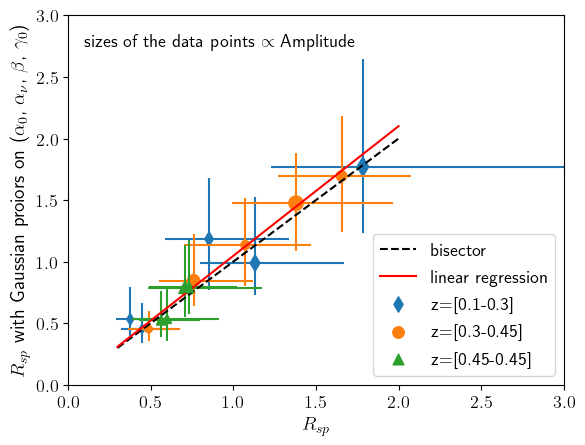}
\caption{\label{linearRel} Linear regression between the splashback radius computed assuming Gaussian priors on the parameters that describe the Einasto profile and the transition term and the one computed using our reference case with fixed values. The red line has been computed using the \texttt{ODR} routine in \texttt{Python} considering the uncertainties in both axes.}
\end{figure}

\begin{figure*}
\centering
\includegraphics[width=\hsize]{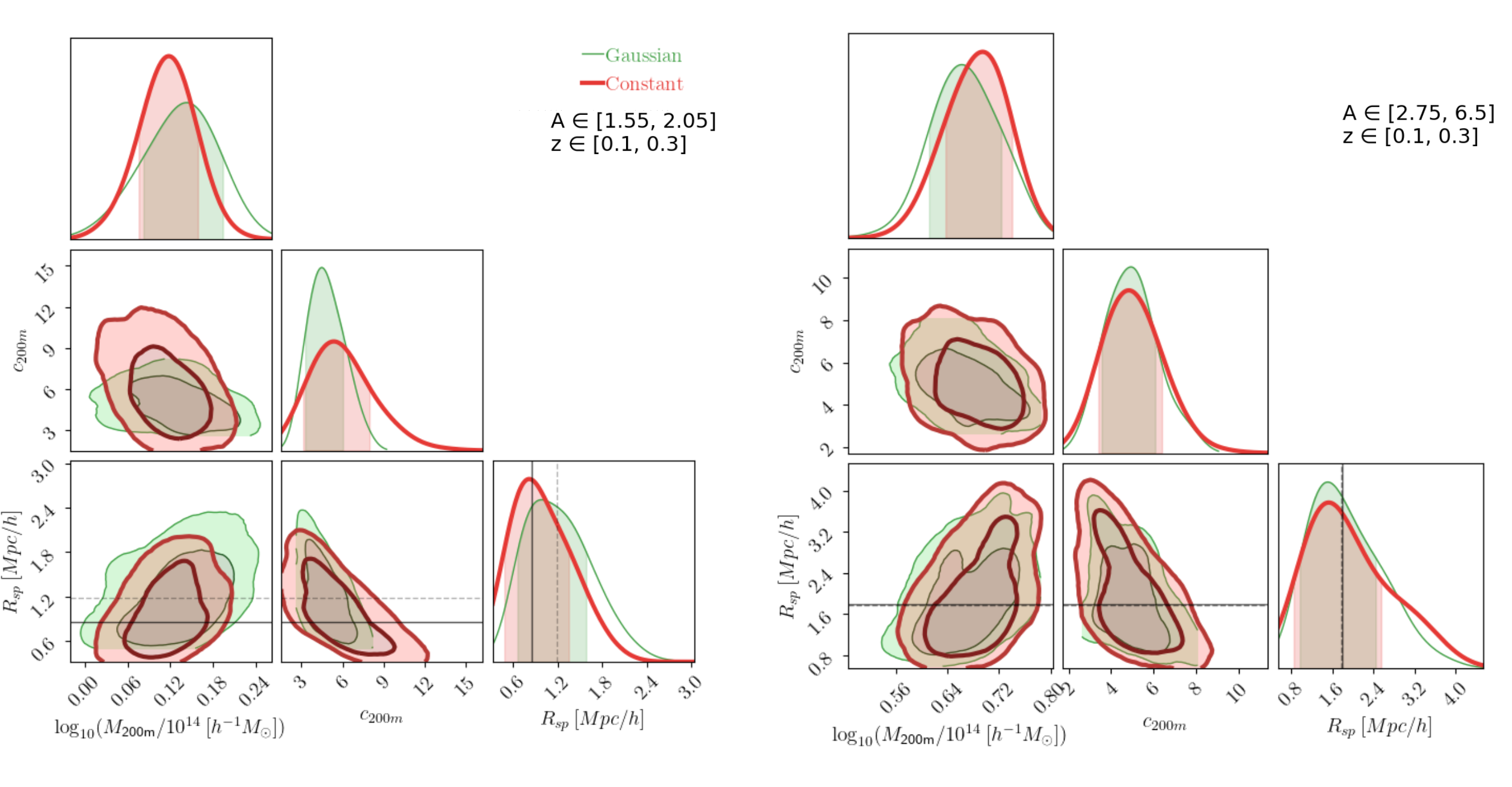}
\caption{\label{new_posteriors}Posterior distributions of the recovered mass, concentration, and splashback radius when assuming constant (red) and Gaussian (green) priors for the parameters $\alpha_0$, $\alpha_{\nu}$, $\beta$ and $\gamma_0$ that
describe the Einasto profile and the transition term in the DK14 model. Left and right panels show two different amplitude bins in the first redshift interval. The left refers to the third blue data point in Fig. \ref{linearRel} where the two splashback radii differ most, while the right one refers to to the largest amplitude bin (corresponding to the last blue data point in Fig. \ref{linearRel}),  as also presented in Fig. \ref{figDKposteriorsRsp}.}
\end{figure*}

However, it is worth underlining that assuming a constant value for the parameters that express the terms $\alpha$, $\beta$, and $\gamma$ in the Einasto profile (Eq. \ref{dk14}) and in the transition term (Eq. \ref{trans}) could be conservative and accurate only for a mass-selected cluster sample. In order to understand how the priors on the parameters impact the splashback radius characterisation, we have run another 
sample of Monte Carlo Markov Chains assuming Gaussian priors with a standard deviation of 10\%
with mean values $\alpha_0=0.155$ and $\alpha_{\nu}=0.0096$  ($\alpha = \alpha_0 + \alpha_{\nu} \nu_{vir}^2$) \citep{gao08}, $\beta=4$ and $\gamma_0=4$ ($\gamma=\gamma_0\nu_{vir}$). In Fig. \ref{linearRel}, we show the scaling relation between the two recovered values of the splashback radius, colouring the data points with respect to the different considered redshift bins. We can notice that within the 1 $\sigma$ credibility region, the two estimates are consistent with each other. The red line represents the linear relation, computed using the \texttt{ODR} routine in \texttt{Python} accounting for the uncertainties in both axes, that has the following parameters:
\begin{equation}
R_{\rm sp,\,Gaussian} =  (1.05 \pm 0.09 )R_{\rm sp} + ( -0.006 \pm 0.062).
\end{equation}
From the figure, we can notice that the third blue data point referring to the third amplitude bin in the first redshift interval presents the largest deviation. For this stacked clusters' sample, in the left panel of Fig. \ref{new_posteriors}, we show the posterior distributions of the recovered mass, concentration, and splashback radius for the two cases when assuming constant and Gaussian priors for the parameters  $\alpha_0$, $\alpha_{\nu}$, $\beta$, and $\gamma_0$. We can see that for the Gaussian case, not only $R_{\rm sp}$ tends to be slightly larger, but also the mass moves toward larger values while the concentration diminishes. In the right panel, we show the posteriors for the largest amplitude bin of the first redshift interval, as also shown in Fig. \ref{figDKposteriorsRsp}. For this clusters' sample, all recovered parameters are close to each other in both cases. The green dashed curves in Fig. \ref{figDlogrho} show the logarithmic derivative of the recovered parameters when assuming Gaussian priors for the parameters that enter in the Einasto profile and the transition term in the DK14. We notice that this choice does not particularly impact our results, considering the uncertainties.
With future data, we expect to better understand the cluster selection effects that propagate in the parameters of the Einasto profile and the transition term. We will discuss our final results considering the model choice with a constant parametrisation of $\alpha$, $\beta$ and $\gamma$.

Different interpretations of the splashback radius have been given in recent years using numerical simulations and observational data. In particular, \citet{more15} have correlated the splashback radius in simulations with the peak heights $\nu_{200m}$ of the primordial density field in which cluster-size haloes are embedded \citep{bardeen86,sheth99b,sheth01b,giocoli07} and with the accretion rate.

\begin{figure}
\centering
\includegraphics[width=\hsize]{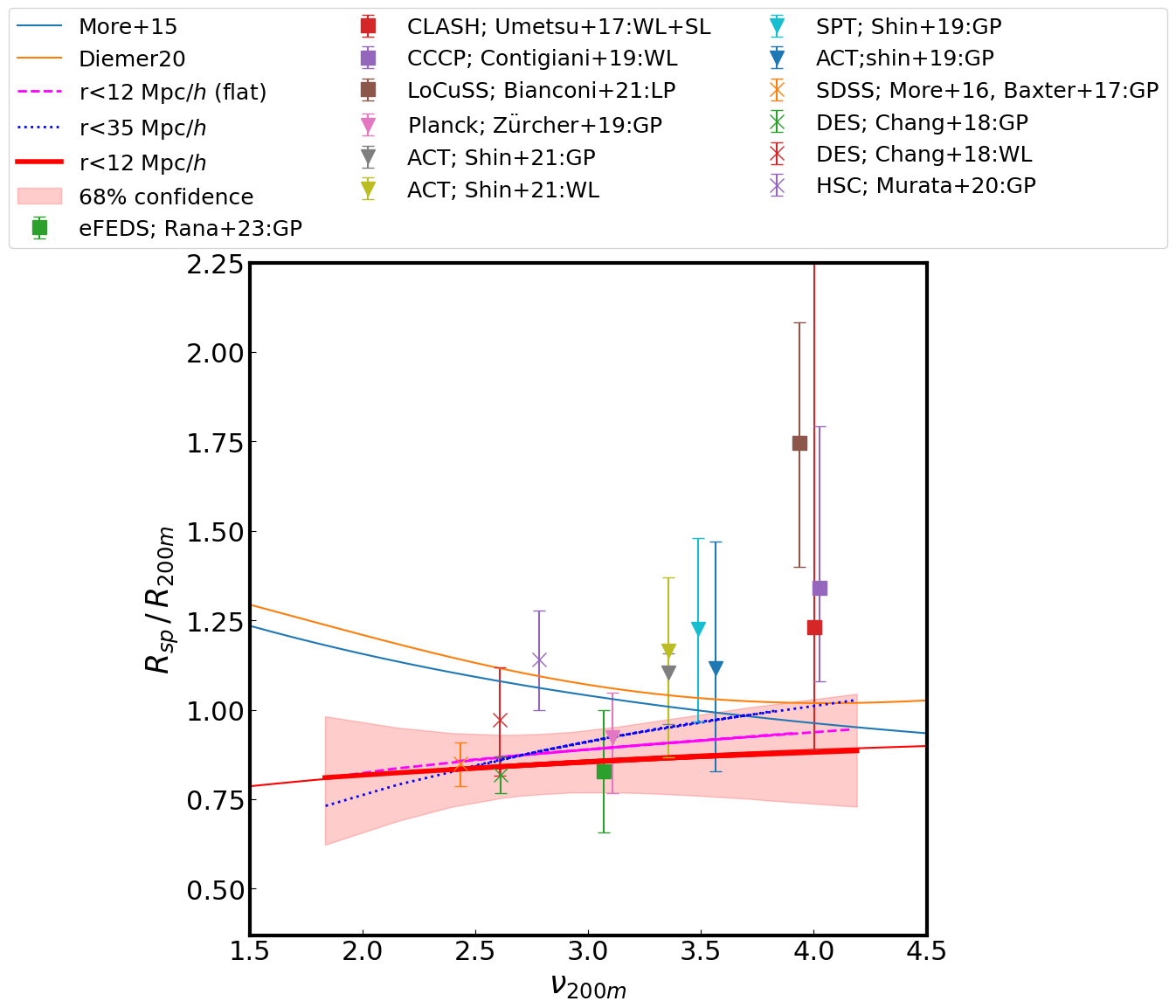}
\caption{\label{figfinal} Rescaled splashback radius as a function of the cluster peak height. The red solid curve and the corresponding shaded region represent the result of our reference fitting procedure model. 
The blue and orange curves correspond to the model predictions by \citet{more15} and \citet{diemer20}, respectively. 
The different data points represent the relation measured by different authors, using various selected galaxy cluster samples and observables: galaxy projected correlation function (GP), weak lensing (WL), and in combination of weak and strong lensing (WL+SL) and stacked luminosity profiles (LP). X-ray, SZ and optically selected clusters are displayed using squares, triangles and crosses.}
\end{figure}

To compare to this theoretical finding, we collect our results and measurements from various observational analyses in Fig.\ref{figfinal}, where we plot $R_{sp}/R_{200m}$ versus $\nu_{200m}$. The solid blue and orange curves represent the recent theoretical models presented by \citet{more15} and \citet{diemer20}, respectively. 
The data points of different colours and shapes indicate cluster samples selected using different methods: filled squares indicate systems selected via X-ray data, triangles via the Sunyaev–Zeldovich (SZ) effect, and crosses mark optically selected clusters.  
In addition,  the suffixes indicate different observables adopted to define the splashback radius: GP stands for galaxy projected correlation function, WL for weak lensing, WL+SL for the combination of weak and strong lensing, and LP when using the stacked luminosity profile of satellite galaxies. 
On average, X-ray clusters tend to have larger peak heights than the SZ and the optically selected ones because of the different instrumental sensitivity that translates into a diverse selection function.

The red solid line displays our findings, modelled by the relation:
\begin{equation}
\dfrac{R_{sp}}{R_{200m}} = A \left( 1 + B e^{-\nu_{200m}/2.44}\right)\,,
\end{equation}
as proposed by \citet{more15}. 
Combining all redshifts -- being the relation as a function of the peak-height $\nu_{200m}$ independent of redshift, the median, and $16th$ and $84th$ percentiles of the distributions from our MCMC analyses can be read as: 
\begin{equation}
A=0.95^{+0.32}_{-0.34} ,\qquad 
B=-0.31^{+1.5}_{-0.73}.
\end{equation}
For both parameters, we assume wide flat priors between -20 and 20. For comparison purposes, in the figure, dashed magenta and dotted blue lines show the results when assuming flat priors for $\log_{10}\left(M_{200m}\right)$ and $c_{200m}$, and when considering Gaussian priors but the data points up to 35 Mpc/$h$, respectively. Those results highlight that the choice of priors has a minor impact on the splashback radius peak height relation than the adopted radial range extension.

The trend of the data points in Fig.~\ref{figfinal} indicates possible biases in characterising the splashback radius due to cluster selection effects, with respect to a pure mass-selected sample -- as done in simulations. Clusters selected via optical data, based on their member galaxies and richness properties, tend to be a biased sample of the average galaxy cluster population.
Probably, optical selection favours the identification of systems that are prolate and oriented along the line-of-sight. 
Those clusters show a splashback radius that is approximately $20\%$ smaller than the one expected from numerical simulations \citep{more16,baxter17,chang18}. 
SZ-selected clusters are less biased and tend to be more consistent with the prediction from numerical simulation models. 
X-ray clusters appear more biased from the other side, on average, more extended -- probably oblate in the plane of the sky and possess a splashback radius outside $R_{200}$. However, the figure shows that observational-derived radii present some tension with the predictions based on mass-selected clusters from numerical simulations. A more detailed analysis of the selection function could be the appropriate path to investigate that we plan to perform in a future dedicated study, also using the \texttt{AMICO} cluster finder on simulated photometric data.
 
\begin{table}
\begin{center}    
\caption{Adopted prior parameters for the DK14 model when modelling 
the cluster+lss data set is used to constrain the splashback radius.}
\label{tabpars_clusterslss}
\begin{tabular}{lc}
\hline
$\log_{10} \left( M_{\rm 200m} / h^{-1} M_{\odot}\right)$ & Gaussian$^a$  \\
$c_{\rm 200m}$ & Gaussian$^a$ \\
$r_{{\rm t}}/R_{\rm 200m}$ & $[0.8,5]$ \\ 
$b_{\rm e}$ & $[0.1,4]$  \\ 
$s_{\rm e}$ & $[0.5,2]$  \\ 
 \hline 
\end{tabular}
\end{center}
\footnotesize{
$^{a}$ as derived from the cluster data model.}
\end{table}

\section{Summary and conclusions}
\label{sec.summary}

In this work, we use weak gravitational lensing data to characterize the cluster splashback radii of optically selected clusters in the KiDS-DR3 data. 
We exploit photometric redshifts with corresponding probability distribution functions \citep{kuijken15, dejong17}, a global improved photometric calibration with respect to previous releases, weak-lensing shear catalogues
\citep{kuijken15, hildebrandt17} and lensing-optimised image data.
Our cluster sample, used to build the stacked shear profiles up to the cluster outskirts was obtained using the \texttt{AMICO} algorithm \citep{bellagamba18}. 
The redshift-amplitude binning of the cluster sample has been constructed as in \citet{bellagamba19,giocoli21,ingoglia22,lesci22}.
We model the excess surface mass density of the stacked signal using the \citet{diemer14} profile that is able to discriminate the orbiting from the infalling component and characterises the scale where the slope of the density profile steepens.  
Our work supports the use of the stacking method in weak lensing to calibrate galaxy cluster scaling relations and shows full agreement with previous results when using a different profile (the BMO one) as the reference model.

We constructed two data sets from the same cluster and galaxy shear catalogues. 
The first data set has been used to calibrate the scaling relations, namely in halo model formalism the $1h$-term, in which we binned the stacked radial excess surface mass density profiles in $14$ intervals from 0.1 to 3.16 Mpc/$h$ as in \citet{bellagamba19,giocoli21}; while the second one has been considered to account for the contribution of the signal on much larger separation from the cluster centres, portraying the infalling material, where the analyses have been performed in 30 bins up to 35 Mpc/$h$. 
To keep under control all the observational systematic uncertainties of the survey, we subtracted to these measurements the signal around random centres, randomising \num{10000} times the positions of the $6962$ \texttt{AMICO} clusters.

We neglect in our analysis the mis-centring terms because we demonstrate that they are degenerate with the concentration parameter and negligible when modelling the lensing profile at a very large distance from the cluster centres. 
Modelling the excess surface mass density profiles, we recovered with good precision the splashback radius at the scale where the density profile exhibits a sharp steepening of the slope. 

We find that the splashback radius of our clusters is close to $R_{200m}$, suggesting that our sample could be characterised by selection effects common to optical cluster finders \citep{more15,rana23}. 
This also highlights that optically selected algorithms seem to preferentially select systems that are more compact and with higher density at small scales, possibly oriented along the major axis of the mass tensor ellipsoid \citep{wu22}.
A future work will be dedicated to the study of the effects of the cluster selection function in the splashback radius using dedicated simulations.
 
We expect to reduce the uncertainties, keeping biases and selection effects under control, thanks to the analysis of the new cluster catalogues from the following KiDS data releases and afterwards from the data coming from future wide field surveys like the ESA Euclid mission \citep{euclidredbook,scaramella22}. 
Upcoming deeper galaxy surveys will allow us to detect and stack lensing clusters in smaller redshift bins and to extend the analysis up to higher redshifts, enabling us to break parameter degeneracies and limit systematic uncertainties \citep{oguri11}. 
Thanks to the multi-wavelength follow up it will be possible to investigate the cluster selection effects in more detail.

The significance of the stacked weak lensing signal can be incremented by considering a larger number of clusters, a more numerous sample of background sources (deeper surveys), and a wider survey area in order to better characterise the splashback radius. 
These aspects will be reached by future wide-field surveys, which will provide a significantly improved data set compared to the one used in this work. The Euclid-wide survey \citep{scaramella22}, for example, is expected to cover an area of about \num{15000} deg$^2$ and it will detect approximately $30$ galaxies per square arc-minute, with a median redshift larger than $0.9$. 
In addition, it will also give us the possibility to divide the sample in terms of accretion rates and environments and study further effects that influence the splashback scale of galaxy clusters. 

We conclude by underling that our methodology of finding clusters -- with AMICO -- and our statistical analyses and models in designing the weak lensing shear profiles around them, discussed here, will represent a reference starting point and a milestone for cluster cosmology in near future experiments such as Euclid \citep{sartoris16,adam19}.

\section*{Acknowledgements}
We thank the anonymous reviewer for the useful comments and suggestions that helped improve the presentation of our results. 
We also thank Joachim Harnois-Déraps for carefully reading our manuscript and for his fruitful comments. 
CG and GD thank Benedikt Diemer for our nice discussions in Garching during the Miapbp meeting 2022: ADVANCES IN COSMOLOGY THROUGH NUMERICAL SIMULATIONS.
GC thanks the support from INAF theory Grant 2022: Illuminating Dark Matter using Weak Lensing by Cluster Satellites, PI: Carlo Giocoli. LM acknowledges the grants ASI n.I/023/12/0 and ASI n.2018- 23-HH.0. GD acknowledges the funding by
the European Union - $NextGenerationEU$, in the framework of the HPC project – “National Centre for HPC, Big Data and Quantum Computing” (PNRR - M4C2 - I1.4 - CN00000013 – CUP J33C22001170001).
GC, LM, FM  and MS acknowledge the financial contribution from the grant
PRIN-MUR 2022 20227RNLY3 “The concordance cosmological model: stress-tests with galaxy clusters” supported by Next Generation EU.
MS acknowledges financial contributions from contract ASI-INAF n.2017-14-H.0, contract INAF mainstream project 1.05.01.86.10, and INAF Theory Grant 2023: Gravitational lensing detection of matter distribution at galaxy cluster boundaries and beyond (1.05.23.06.17).
\bibliography{main}

\begin{thebibliography}{119}
\expandafter\ifx\csname natexlab\endcsname\relax\def\natexlab#1{#1}\fi

\bibitem[{{Adhikari} {et~al.}(2014){Adhikari}, {Dalal}, \&
  {Chamberlain}}]{adhikari14}
{Adhikari}, S., {Dalal}, N., \& {Chamberlain}, R.~T. 2014,
  \href{http://dx.doi.org/10.1088/1475-7516/2014/11/019}{\color{magenta}\jcap},
  \href{https://ui.adsabs.harvard.edu/abs/2014JCAP...11..019A}{2014, 019}

\bibitem[{{Adhikari} {et~al.}(2018){Adhikari}, {Sakstein}, {Jain}, {Dalal}, \&
  {Li}}]{adhikari18}
{Adhikari}, S., {Sakstein}, J., {Jain}, B., {Dalal}, N., \& {Li}, B. 2018,
  \href{http://dx.doi.org/10.1088/1475-7516/2018/11/033}{\color{magenta}\jcap},
  \href{https://ui.adsabs.harvard.edu/abs/2018JCAP...11..033A}{2018, 033}

\bibitem[{{Adhikari} {et~al.}(2021){Adhikari}, {Shin}, {Jain}, {Hilton},
  {Baxter}, {Chang}, {Wechsler}, {Battaglia}, {Bond}, {Bocquet}, {Choi},
  {DeRose}, {Devlin}, {Dunkley}, {Evrard}, {Ferraro}, {Hill}, {Hughes},
  {Gallardo}, {Lokken}, {MacInnis}, {Madhavacheril}, {McMahon}, {Nati},
  {Newburgh}, {Niemack}, {Page}, {Palmese}, {Partridge}, {Rozo}, {Rykoff},
  {Salatino}, {Schillaci}, {Sehgal}, {Sif{\'o}n}, {To}, {Wollack}, {Wu}, {Xu},
  {Aguena}, {Allam}, {Amon}, {Annis}, {Avila}, {Bacon}, {Bertin}, {Bhargava},
  {Brooks}, {Burke}, {Rosell}, {Kind}, {Carretero}, {Castander}, {Choi},
  {Costanzi}, {da Costa}, {Vicente}, {Desai}, {Diehl}, {Doel}, {Everett},
  {Ferrero}, {Fert{\'e}}, {Flaugher}, {Fosalba}, {Frieman},
  {Garc{\'\i}a-Bellido}, {Gaztanaga}, {Gruen}, {Gruendl}, {Gschwend},
  {Gutierrez}, {Hartley}, {Hinton}, {Hollowood}, {Honscheid}, {James},
  {Jeltema}, {Kuehn}, {Kuropatkin}, {Lahav}, {Lima}, {Maia}, {Marshall},
  {Martini}, {Melchior}, {Menanteau}, {Miquel}, {Morgan}, {L.~C. Ogando},
  {Paz-Chinch{\'o}n}, {Malag{\'o}n}, {Sanchez}, {Santiago}, {Scarpine},
  {Serrano}, {Sevilla-Noarbe}, {Smith}, {Soares-Santos}, {Suchyta}, {E.~C.
  Swanson}, {Varga}, {Wilkinson}, {Zhang}, {Austermann}, {Beall}, {Becker},
  {Denison}, {Duff}, {Hilton}, {Hubmayr}, {Ullom}, {Lanen}, {Vale}, {Vale}, \&
  {Vale}}]{adhikari21}
{Adhikari}, S., {Shin}, T.-h., {Jain}, B., {et~al.} 2021,
  \href{http://dx.doi.org/10.3847/1538-4357/ac0bbc}{\color{magenta}\apj},
  \href{https://ui.adsabs.harvard.edu/abs/2021ApJ...923...37A}{923, 37}

\bibitem[{{Angulo} {et~al.}(2009){Angulo}, {Lacey}, {Baugh}, \&
  {Frenk}}]{angulo09}
{Angulo}, R.~E., {Lacey}, C.~G., {Baugh}, C.~M., \& {Frenk}, C.~S. 2009,
  \href{http://dx.doi.org/10.1111/j.1365-2966.2009.15333.x}{\color{magenta}\mnras},
  \href{http://adsabs.harvard.edu/abs/2009MNRAS.399..983A}{399, 983}

\bibitem[{Angulo {et~al.}(2012)Angulo, Springel, White, Jenkins, Baugh, \&
  Frenk}]{angulo12}
Angulo, R.~E., Springel, V., White, S. D.~M., {et~al.} 2012,
  \href{http://dx.doi.org/10.1111/j.1365-2966.2012.21830.x}{\color{magenta}\mnras},
  426, 426

\bibitem[{{Asgari} {et~al.}(2021){Asgari}, {Lin}, {Joachimi}, {Giblin},
  {Heymans}, {Hildebrandt}, {Kannawadi}, {St{\"o}lzner}, {Tr{\"o}ster}, {van
  den Busch}, {Wright}, {Bilicki}, {Blake}, {de Jong}, {Dvornik}, {Erben},
  {Getman}, {Hoekstra}, {K{\"o}hlinger}, {Kuijken}, {Miller}, {Radovich},
  {Schneider}, {Shan}, \& {Valentijn}}]{asgari21}
{Asgari}, M., {Lin}, C.-A., {Joachimi}, B., {et~al.} 2021,
  \href{http://dx.doi.org/10.1051/0004-6361/202039070}{\color{magenta}\aap},
  \href{https://ui.adsabs.harvard.edu/abs/2021A&A...645A.104A}{645, A104}

\bibitem[{{Baltz} {et~al.}(2009){Baltz}, {Marshall}, \& {Oguri}}]{baltz09}
{Baltz}, E.~A., {Marshall}, P., \& {Oguri}, M. 2009,
  \href{http://dx.doi.org/10.1088/1475-7516/2009/01/015}{\color{magenta}\jcap},
  \href{https://ui.adsabs.harvard.edu/abs/2009JCAP...01..015B}{2009, 015}

\bibitem[{{Bardeen} {et~al.}(1986){Bardeen}, {Bond}, {Kaiser}, \&
  {Szalay}}]{bardeen86}
{Bardeen}, J.~M., {Bond}, J.~R., {Kaiser}, N., \& {Szalay}, A.~S. 1986,
  \href{http://dx.doi.org/10.1086/164143}{\color{magenta}\apj},
  \href{http://adsabs.harvard.edu/abs/1986ApJ...304...15B}{304, 15}

\bibitem[{{Bartelmann} \& {Schneider}(2001)}]{bartelmann01}
{Bartelmann}, M. \& {Schneider}, P. 2001,
  \href{http://dx.doi.org/10.1016/S0370-1573(00)00082-X}{\color{magenta}Physics
  Reports}, \href{http://adsabs.harvard.edu/abs/2001PhR...340..291B}{340, 291}

\bibitem[{{Baxter} {et~al.}(2017){Baxter}, {Chang}, {Jain}, {Adhikari},
  {Dalal}, {Kravtsov}, {More}, {Rozo}, {Rykoff}, \& {Sheth}}]{baxter17}
{Baxter}, E., {Chang}, C., {Jain}, B., {et~al.} 2017,
  \href{http://dx.doi.org/10.3847/1538-4357/aa6ff0}{\color{magenta}\apj},
  \href{https://ui.adsabs.harvard.edu/abs/2017ApJ...841...18B}{841, 18}

\bibitem[{{Bellagamba} {et~al.}(2018){Bellagamba}, {Roncarelli}, {Maturi}, \&
  {Moscardini}}]{bellagamba18}
{Bellagamba}, F., {Roncarelli}, M., {Maturi}, M., \& {Moscardini}, L. 2018,
  \href{http://dx.doi.org/10.1093/mnras/stx2701}{\color{magenta}\mnras},
  \href{https://ui.adsabs.harvard.edu/abs/2018MNRAS.473.5221B}{473, 5221}

\bibitem[{{Bellagamba} {et~al.}(2019){Bellagamba}, {Sereno}, {Roncarelli},
  {Maturi}, {Radovich}, {Bardelli}, {Puddu}, {Moscardini}, {Getman},
  {Hildebrandt}, \& {Napolitano}}]{bellagamba19}
{Bellagamba}, F., {Sereno}, M., {Roncarelli}, M., {et~al.} 2019,
  \href{http://dx.doi.org/10.1093/mnras/stz090}{\color{magenta}\mnras},
  \href{https://ui.adsabs.harvard.edu/abs/2019MNRAS.484.1598B}{484, 1598}

\bibitem[{{Ben{\'\i}tez}(2000)}]{benitez00}
{Ben{\'\i}tez}, N. 2000,
  \href{http://dx.doi.org/10.1086/308947}{\color{magenta}\apj},
  \href{https://ui.adsabs.harvard.edu/abs/2000ApJ...536..571B}{536, 571}

\bibitem[{{Bonafede} {et~al.}(2011){Bonafede}, {Dolag}, {Stasyszyn}, {Murante},
  \& {Borgani}}]{bonafede11}
{Bonafede}, A., {Dolag}, K., {Stasyszyn}, F., {Murante}, G., \& {Borgani}, S.
  2011,
  \href{http://dx.doi.org/10.1111/j.1365-2966.2011.19523.x}{\color{magenta}\mnras},
  \href{http://adsabs.harvard.edu/abs/2011MNRAS.418.2234B}{418, 2234}

\bibitem[{{Bridle} \& {King}(2007)}]{bridle07}
{Bridle}, S. \& {King}, L. 2007,
  \href{http://dx.doi.org/10.1088/1367-2630/9/12/444}{\color{magenta}New
  Journal of Physics},
  \href{https://ui.adsabs.harvard.edu/abs/2007NJPh....9..444B}{9, 444}

\bibitem[{{Bryan} \& {Norman}(1998)}]{bryan98}
{Bryan}, G.~L. \& {Norman}, M.~L. 1998,
  \href{http://dx.doi.org/10.1086/305262}{\color{magenta}\apj},
  \href{http://adsabs.harvard.edu/abs/1998ApJ...495...80B}{495, 80}

\bibitem[{{Bullock} {et~al.}(2001){Bullock}, {Kolatt}, {Sigad}, {Somerville},
  {Kravtsov}, {Klypin}, {Primack}, \& {Dekel}}]{bullock01a}
{Bullock}, J.~S., {Kolatt}, T.~S., {Sigad}, Y., {et~al.} 2001, \mnras,
  \href{http://adsabs.harvard.edu/abs/2001MNRAS.321..559B}{321, 559}

\bibitem[{{Cacciato} {et~al.}(2012){Cacciato}, {Lahav}, {van den Bosch},
  {Hoekstra}, \& {Dekel}}]{cacciato12}
{Cacciato}, M., {Lahav}, O., {van den Bosch}, F.~C., {Hoekstra}, H., \&
  {Dekel}, A. 2012,
  \href{http://dx.doi.org/10.1111/j.1365-2966.2012.21762.x}{\color{magenta}\mnras},
  \href{http://adsabs.harvard.edu/abs/2012MNRAS.426..566C}{426, 566}

\bibitem[{{Cacciato} {et~al.}(2009){Cacciato}, {van den Bosch}, {More}, {Li},
  {Mo}, \& {Yang}}]{cacciato09}
{Cacciato}, M., {van den Bosch}, F.~C., {More}, S., {et~al.} 2009,
  \href{http://dx.doi.org/10.1111/j.1365-2966.2008.14362.x}{\color{magenta}\mnras},
  \href{http://adsabs.harvard.edu/abs/2009MNRAS.394..929C}{394, 929}

\bibitem[{{Capaccioli} \& {Schipani}(2011)}]{capaccioli11}
{Capaccioli}, M. \& {Schipani}, P. 2011, The Messenger,
  \href{https://ui.adsabs.harvard.edu/abs/2011Msngr.146....2C}{146, 2}

\bibitem[{{Chang} {et~al.}(2018){Chang}, {Baxter}, {Jain}, {S{\'a}nchez},
  {Adhikari}, {Varga}, {Fang}, {Rozo}, {Rykoff}, {Kravtsov}, {Gruen},
  {Hartley}, {Huff}, {Jarvis}, {Kim}, {Prat}, {MacCrann}, {McClintock},
  {Palmese}, {Rapetti}, {Rollins}, {Samuroff}, {Sheldon}, {Troxel}, {Wechsler},
  {Zhang}, {Zuntz}, {Abbott}, {Abdalla}, {Allam}, {Annis}, {Bechtol},
  {Benoit-L{\'e}vy}, {Bernstein}, {Brooks}, {Buckley-Geer}, {Carnero Rosell},
  {Carrasco Kind}, {Carretero}, {D'Andrea}, {da Costa}, {Davis}, {Desai},
  {Diehl}, {Dietrich}, {Drlica-Wagner}, {Eifler}, {Flaugher}, {Fosalba},
  {Frieman}, {Garc{\'\i}a-Bellido}, {Gaztanaga}, {Gerdes}, {Gruendl},
  {Gschwend}, {Gutierrez}, {Honscheid}, {James}, {Jeltema}, {Krause}, {Kuehn},
  {Lahav}, {Lima}, {March}, {Marshall}, {Martini}, {Melchior}, {Menanteau},
  {Miquel}, {Mohr}, {Nord}, {Ogando}, {Plazas}, {Sanchez}, {Scarpine},
  {Schindler}, {Schubnell}, {Sevilla-Noarbe}, {Smith}, {Smith},
  {Soares-Santos}, {Sobreira}, {Suchyta}, {Swanson}, {Tarle}, {Weller}, \& {DES
  Collaboration}}]{chang18}
{Chang}, C., {Baxter}, E., {Jain}, B., {et~al.} 2018,
  \href{http://dx.doi.org/10.3847/1538-4357/aad5e7}{\color{magenta}\apj},
  \href{https://ui.adsabs.harvard.edu/abs/2018ApJ...864...83C}{864, 83}

\bibitem[{{Chisari} {et~al.}(2014){Chisari}, {Mandelbaum}, {Strauss}, {Huff},
  \& {Bahcall}}]{chisari14}
{Chisari}, N.~E., {Mandelbaum}, R., {Strauss}, M.~A., {Huff}, E.~M., \&
  {Bahcall}, N.~A. 2014,
  \href{http://dx.doi.org/10.1093/mnras/stu1786}{\color{magenta}\mnras},
  \href{https://ui.adsabs.harvard.edu/abs/2014MNRAS.445..726C}{445, 726}

\bibitem[{{Contigiani} {et~al.}(2023){Contigiani}, {Hoekstra}, {Brouwer},
  {Dvornik}, {Fortuna}, {Sif{\'o}n}, {Yan}, \& {Vakili}}]{contigiani23}
{Contigiani}, O., {Hoekstra}, H., {Brouwer}, M.~M., {et~al.} 2023,
  \href{http://dx.doi.org/10.1093/mnras/stac3027}{\color{magenta}\mnras},
  \href{https://ui.adsabs.harvard.edu/abs/2023MNRAS.518.2640C}{518, 2640}

\bibitem[{{Contigiani} {et~al.}(2019a){Contigiani}, {Vardanyan}, \&
  {Silvestri}}]{contigiani19a}
{Contigiani}, O., {Vardanyan}, V., \& {Silvestri}, A. 2019a,
  \href{http://dx.doi.org/10.1103/PhysRevD.99.064030}{\color{magenta}\prd},
  \href{https://ui.adsabs.harvard.edu/abs/2019PhRvD..99f4030C}{99, 064030}

\bibitem[{{Crocce} {et~al.}(2010){Crocce}, {Fosalba}, {Castander}, \&
  {Gazta{\~n}aga}}]{crocce10}
{Crocce}, M., {Fosalba}, P., {Castander}, F.~J., \& {Gazta{\~n}aga}, E. 2010,
  \href{http://dx.doi.org/10.1111/j.1365-2966.2009.16194.x}{\color{magenta}\mnras},
  \href{http://adsabs.harvard.edu/abs/2010MNRAS.403.1353C}{403, 1353}

\bibitem[{{Cui} {et~al.}(2018){Cui}, {Knebe}, {Yepes}, {Pearce}, {Power},
  {Dave}, {Arth}, {Borgani}, {Dolag}, {Elahi}, {Mostoghiu}, {Murante}, {Rasia},
  {Stoppacher}, {Vega-Ferrero}, {Wang}, {Yang}, {Benson}, {Cora}, {Croton},
  {Sinha}, {Stevens}, {Vega-Mart{\'\i}nez}, {Arthur}, {Baldi}, {Ca{\~n}as},
  {Cialone}, {Cunnama}, {De Petris}, {Durando}, {Ettori}, {Gottl{\"o}ber},
  {Nuza}, {Old}, {Pilipenko}, {Sorce}, \& {Welker}}]{cui18}
{Cui}, W., {Knebe}, A., {Yepes}, G., {et~al.} 2018,
  \href{http://dx.doi.org/10.1093/mnras/sty2111}{\color{magenta}\mnras},
  \href{https://ui.adsabs.harvard.edu/abs/2018MNRAS.480.2898C}{480, 2898}

\bibitem[{{de Jong} {et~al.}(2017){de Jong}, {Verdoes Kleijn}, {Erben},
  {Hildebrandt}, {Kuijken}, {Sikkema}, {Brescia}, \& {et al.}}]{dejong17}
{de Jong}, J. T.~A., {Verdoes Kleijn}, G.~A., {Erben}, T., {et~al.} 2017,
  \href{http://dx.doi.org/10.1051/0004-6361/201730747}{\color{magenta}\aap},
  \href{https://ui.adsabs.harvard.edu/abs/2017A&A...604A.134D}{604, A134}

\bibitem[{{de Jong} {et~al.}(2013){de Jong}, {Verdoes Kleijn}, {Kuijken}, \&
  {Valentijn}}]{dejong13}
{de Jong}, J. T.~A., {Verdoes Kleijn}, G.~A., {Kuijken}, K.~H., \& {Valentijn},
  E.~A. 2013,
  \href{http://dx.doi.org/10.1007/s10686-012-9306-1}{\color{magenta}Experimental
  Astronomy}, \href{https://ui.adsabs.harvard.edu/abs/2013ExA....35...25D}{35,
  25}

\bibitem[{{Despali} {et~al.}(2016){Despali}, {Giocoli}, {Angulo}, {Tormen},
  {Sheth}, {Baso}, \& {Moscardini}}]{despali16}
{Despali}, G., {Giocoli}, C., {Angulo}, R.~E., {et~al.} 2016,
  \href{http://dx.doi.org/10.1093/mnras/stv2842}{\color{magenta}\mnras},
  \href{http://adsabs.harvard.edu/abs/2016MNRAS.456.2486D}{456, 2486}

\bibitem[{{Despali} {et~al.}(2020){Despali}, {Lovell}, {Vegetti}, {Crain}, \&
  {Oppenheimer}}]{despali20}
{Despali}, G., {Lovell}, M., {Vegetti}, S., {Crain}, R.~A., \& {Oppenheimer},
  B.~D. 2020,
  \href{http://dx.doi.org/10.1093/mnras/stz3068}{\color{magenta}\mnras},
  \href{https://ui.adsabs.harvard.edu/abs/2020MNRAS.491.1295D}{491, 1295}

\bibitem[{{Despali} {et~al.}(2022){Despali}, {Walls}, {Vegetti}, {Sparre},
  {Vogelsberger}, \& {Zavala}}]{despali22}
{Despali}, G., {Walls}, L.~G., {Vegetti}, S., {et~al.} 2022,
  \href{http://dx.doi.org/10.1093/mnras/stac2521}{\color{magenta}\mnras},
  \href{https://ui.adsabs.harvard.edu/abs/2022MNRAS.516.4543D}{516, 4543}

\bibitem[{{Diemer}(2017)}]{diemer17}
{Diemer}, B. 2017,
  \href{http://dx.doi.org/10.3847/1538-4365/aa799c}{\color{magenta}\apjs},
  \href{https://ui.adsabs.harvard.edu/abs/2017ApJS..231....5D}{231, 5}

\bibitem[{{Diemer}(2018)}]{diemer18}
{Diemer}, B. 2018,
  \href{http://dx.doi.org/10.3847/1538-4365/aaee8c}{\color{magenta}\apjs},
  \href{https://ui.adsabs.harvard.edu/abs/2018ApJS..239...35D}{239, 35}

\bibitem[{{Diemer}(2020)}]{diemer20}
{Diemer}, B. 2020,
  \href{http://dx.doi.org/10.3847/1538-4365/abbf51}{\color{magenta}\apjs},
  \href{https://ui.adsabs.harvard.edu/abs/2020ApJS..251...17D}{251, 17}

\bibitem[{{Diemer} \& {Joyce}(2019)}]{diemer19}
{Diemer}, B. \& {Joyce}, M. 2019,
  \href{http://dx.doi.org/10.3847/1538-4357/aafad6}{\color{magenta}\apj},
  \href{https://ui.adsabs.harvard.edu/abs/2019ApJ...871..168D}{871, 168}

\bibitem[{{Diemer} \& {Kravtsov}(2014)}]{diemer14}
{Diemer}, B. \& {Kravtsov}, A.~V. 2014,
  \href{http://dx.doi.org/10.1088/0004-637X/789/1/1}{\color{magenta}\apj},
  \href{https://ui.adsabs.harvard.edu/abs/2014ApJ...789....1D}{789, 1}

\bibitem[{{Diemer} {et~al.}(2013){Diemer}, {More}, \& {Kravtsov}}]{diemer13}
{Diemer}, B., {More}, S., \& {Kravtsov}, A.~V. 2013,
  \href{http://dx.doi.org/10.1088/0004-637X/766/1/25}{\color{magenta}\apj},
  \href{http://adsabs.harvard.edu/abs/2013ApJ...766...25D}{766, 25}

\bibitem[{{Einasto}(1965)}]{einasto65}
{Einasto}, J. 1965, Trudy Astrofizicheskogo Instituta Alma-Ata,
  \href{http://adsabs.harvard.edu/abs/1965TrAlm...5...87E}{5, 87}

\bibitem[{{Eisenstein} \& {Hu}(1999)}]{eisenstein99}
{Eisenstein}, D.~J. \& {Hu}, W. 1999,
  \href{http://dx.doi.org/10.1086/306640}{\color{magenta}\apj},
  \href{https://ui.adsabs.harvard.edu/abs/1999ApJ...511....5E}{511, 5}

\bibitem[{{Euclid Collaboration} {et~al.}(2024){Euclid Collaboration},
  {Giocoli}, {Meneghetti}, {Rasia}, {Borgani}, {Despali}, {Lesci}, {Marulli},
  {Moscardini}, {Sereno}, {Cui}, {Knebe}, {Yepes}, {Castro}, {Corasaniti},
  {Pires}, {Castignani}, {Schrabback}, {Pratt}, {Le Brun}, {Aghanim},
  {Amendola}, {Auricchio}, {Baldi}, {Bodendorf}, {Bonino}, {Branchini},
  {Brescia}, {Brinchmann}, {Camera}, {Capobianco}, {Carbone}, {Carretero},
  {Castander}, {Castellano}, {Cavuoti}, {Cledassou}, {Congedo}, {Conselice},
  {Conversi}, {Copin}, {Corcione}, {Courbin}, {Cropper}, {Da Silva},
  {Degaudenzi}, {Dinis}, {Dubath}, {Dupac}, {Dusini}, {Farrens}, {Ferriol},
  {Fosalba}, {Frailis}, {Franceschi}, {Fumana}, {Galeotta}, {Garilli},
  {Gillis}, {Grazian}, {Grupp}, {Haugan}, {Holmes}, {Hornstrup}, {Jahnke},
  {K{\"u}mmel}, {Kermiche}, {Kilbinger}, {Kunz}, {Kurki-Suonio}, {Ligori},
  {Lilje}, {Lloro}, {Maiorano}, {Mansutti}, {Marggraf}, {Markovic}, {Massey},
  {Maurogordato}, {Mei}, {Merlin}, {Meylan}, {Moresco}, {Munari}, {Niemi},
  {Nightingale}, {Nutma}, {Padilla}, {Paltani}, {Pasian}, {Pedersen},
  {Pettorino}, {Polenta}, {Poncet}, {Popa}, {Raison}, {Renzi}, {Rhodes},
  {Riccio}, {Romelli}, {Roncarelli}, {Rossetti}, {Saglia}, {Sapone},
  {Sartoris}, {Schneider}, {Secroun}, {Serrano}, {Sirignano}, {Sirri},
  {Stanco}, {Starck}, {Tallada-Cresp{\'\i}}, {Taylor}, {Tereno},
  {Toledo-Moreo}, {Torradeflot}, {Tutusaus}, {Valentijn}, {Valenziano},
  {Vassallo}, {Wang}, {Weller}, {Zamorani}, {Zoubian}, {Andreon}, {Bardelli},
  {Boucaud}, {Bozzo}, {Colodro-Conde}, {Di Ferdinando}, {Fabbian}, {Farina},
  {Israel}, {Keih{\"a}nen}, {Lindholm}, {Mauri}, {Neissner}, {Schirmer},
  {Scottez}, {Tenti}, {Zucca}, {Akrami}, {Baccigalupi}, {Ballardini},
  {Bernardeau}, {Biviano}, {Borlaff}, {Burigana}, {Cabanac}, {Cappi},
  {Carvalho}, {Casas}, {Chambers}, {Cooray}, {Courtois}, {Davini}, {de la
  Torre}, {De Lucia}, {Desprez}, {Dole}, {Escartin}, {Escoffier}, {Ferrero},
  {Finelli}, {Gabarra}, {Ganga}, {Garcia-Bellido}, {George}, {Giacomini},
  {Gozaliasl}, {Hildebrandt}, {Hook}, {Jimenez Mu{\~n}oz}, {Joachimi},
  {Kajava}, {Kansal}, {Kirkpatrick}, {Legrand}, {Loureiro}, {Macias-Perez},
  {Magliocchetti}, {Mainetti}, {Maoli}, {Marcin}, {Martinelli}, {Martinet},
  {Martins}, {Matthew}, {Maurin}, {Metcalf}, {Monaco}, {Morgante}, {Nadathur},
  {Nucita}, {Patrizii}, {Peel}, {Pollack}, {Popa}, {Porciani}, {Potter},
  {P{\"o}ntinen}, {Reimberg}, {S{\'a}nchez}, {Sakr}, {Schneider}, {Sefusatti},
  {Shulevski}, {Spurio Mancini}, {Stadel}, {Steinwagner}, {Valiviita},
  {Veropalumbo}, {Viel}, \& {Zinchenko}}]{giocoli24}
{Euclid Collaboration}, {Giocoli}, C., {Meneghetti}, M., {et~al.} 2024,
  \href{http://dx.doi.org/10.1051/0004-6361/202346058}{\color{magenta}\aap},
  \href{https://ui.adsabs.harvard.edu/abs/2024A&A...681A..67E}{681, A67}

\bibitem[{{Euclid Collaboration: Adam} {et~al.}(2019){Euclid Collaboration:
  Adam}, R., {Vannier}, {Maurogordato}, {Biviano}, {Adami}, {Ascaso},
  {Bellagamba}, {Benoist}, \& {et al.}}]{adam19}
{Euclid Collaboration: Adam}, R., {Vannier}, M., {et~al.} 2019,
  \href{http://dx.doi.org/10.1051/0004-6361/201935088}{\color{magenta}\aap},
  \href{https://ui.adsabs.harvard.edu/abs/2019A&A...627A..23E}{627, A23}

\bibitem[{{Euclid Collaboration: Scaramella} {et~al.}(2022){Euclid
  Collaboration: Scaramella}, R., {Amiaux}, {Mellier}, {Burigana}, {Carvalho},
  {Cuillandre}, {Da Silva}, {Derosa}, {Dinis}, {Maiorano}, {Maris}, {Tereno},
  {Laureijs}, {Boenke}, {Buenadicha}, {Dupac}, {Gaspar Venancio},
  {G{\'o}mez-{\'A}lvarez}, {Hoar}, {Lorenzo Alvarez}, {Racca},
  {Saavedra-Criado}, {Schwartz}, {Vavrek}, {Schirmer}, {Aussel}, {Azzollini},
  {Cardone}, {Cropper}, {Ealet}, {Garilli}, {Gillard}, {Granett}, {Guzzo},
  {Hoekstra}, {Jahnke}, {Kitching}, {Maciaszek}, {Meneghetti}, {Miller},
  {Nakajima}, {Niemi}, {Pasian}, {Percival}, {Pottinger}, {Sauvage},
  {Scodeggio}, {Wachter}, {Zacchei}, {Aghanim}, {Amara}, {Auphan}, {Auricchio},
  {Awan}, {Balestra}, {Bender}, {Bodendorf}, {Bonino}, {Branchini},
  {Brau-Nogue}, {Brescia}, {Candini}, {Capobianco}, {Carbone}, {Carlberg},
  {Carretero}, {Casas}, {Castander}, {Castellano}, {Cavuoti}, {Cimatti},
  {Cledassou}, {Congedo}, {Conselice}, {Conversi}, {Copin}, {Corcione},
  {Costille}, {Courbin}, {Degaudenzi}, {Douspis}, {Dubath}, {Duncan}, {Dusini},
  {Farrens}, {Ferriol}, {Fosalba}, {Fourmanoit}, {Frailis}, {Franceschi},
  {Franzetti}, {Fumana}, {Gillis}, {Giocoli}, {Grazian}, {Grupp}, {Haugan},
  {Holmes}, {Hormuth}, {Hudelot}, {Kermiche}, {Kiessling}, {Kilbinger},
  {Kohley}, {Kubik}, {K{\"u}mmel}, {Kunz}, {Kurki-Suonio}, {Lahav}, {Ligori},
  {Lilje}, {Lloro}, {Mansutti}, {Marggraf}, {Markovic}, {Marulli}, {Massey},
  {Maurogordato}, {Melchior}, {Merlin}, {Meylan}, {Mohr}, {Moresco}, {Morin},
  {Moscardini}, {Munari}, {Nichol}, {Padilla}, {Paltani}, {Peacock},
  {Pedersen}, {Pettorino}, {Pires}, {Poncet}, {Popa}, {Pozzetti}, {Raison},
  {Rebolo}, {Rhodes}, {Rix}, {Roncarelli}, {Rossetti}, {Saglia}, {Schneider},
  {Schrabback}, {Secroun}, {Seidel}, {Serrano}, {Sirignano}, {Sirri},
  {Skottfelt}, {Stanco}, {Starck}, {Tallada-Cresp{\'\i}}, {Tavagnacco},
  {Taylor}, {Teplitz}, {Toledo-Moreo}, {Torradeflot}, {Trifoglio}, {Valentijn},
  {Valenziano}, {Verdoes Kleijn}, {Wang}, {Welikala}, {Weller}, {Wetzstein},
  {Zamorani}, {Zoubian}, {Andreon}, {Baldi}, {Bardelli}, {Boucaud}, {Camera},
  {Di Ferdinando}, {Fabbian}, {Farinelli}, {Galeotta}, {Graci{\'a}-Carpio},
  {Maino}, {Medinaceli}, {Mei}, {Neissner}, {Polenta}, {Renzi}, {Romelli},
  {Rosset}, {Sureau}, {Tenti}, {Vassallo}, {Zucca}, {Baccigalupi},
  {Balaguera-Antol{\'\i}nez}, {Battaglia}, {Biviano}, {Borgani}, {Bozzo},
  {Cabanac}, {Cappi}, {Casas}, {Castignani}, {Colodro-Conde}, {Coupon},
  {Courtois}, {Cuby}, {de la Torre}, {Desai}, {Dole}, {Fabricius}, {Farina},
  {Ferreira}, {Finelli}, {Flose-Reimberg}, {Fotopoulou}, {Ganga}, {Gozaliasl},
  {Hook}, {Keihanen}, {Kirkpatrick}, {Liebing}, {Lindholm}, {Mainetti},
  {Martinelli}, {Martinet}, {Maturi}, {McCracken}, {Metcalf}, {Morgante},
  {Nightingale}, {Nucita}, {Patrizii}, {Potter}, {Riccio}, {S{\'a}nchez},
  {Sapone}, {Schewtschenko}, {Schultheis}, {Scottez}, {Teyssier}, {Tutusaus},
  {Valiviita}, {Viel}, {Vriend}, \& {Whittaker}}]{scaramella22}
{Euclid Collaboration: Scaramella}, R., {Amiaux}, J., {et~al.} 2022,
  \href{http://dx.doi.org/10.1051/0004-6361/202141938}{\color{magenta}\aap},
  \href{https://ui.adsabs.harvard.edu/abs/2022A&A...662A.112E}{662, A112}

\bibitem[{{Fenech Conti} {et~al.}(2017){Fenech Conti}, {Herbonnet}, {Hoekstra},
  {Merten}, {Miller}, \& {Viola}}]{fenechconti17}
{Fenech Conti}, I., {Herbonnet}, R., {Hoekstra}, H., {et~al.} 2017,
  \href{http://dx.doi.org/10.1093/mnras/stx200}{\color{magenta}\mnras},
  \href{https://ui.adsabs.harvard.edu/abs/2017MNRAS.467.1627F}{467, 1627}

\bibitem[{{Fillmore} \& {Goldreich}(1984)}]{fillmore84}
{Fillmore}, J.~A. \& {Goldreich}, P. 1984,
  \href{http://dx.doi.org/10.1086/162070}{\color{magenta}\apj},
  \href{https://ui.adsabs.harvard.edu/abs/1984ApJ...281....1F}{281, 1}

\bibitem[{{Fischbacher} {et~al.}(2023){Fischbacher}, {Kacprzak}, {Blazek}, \&
  {Refregier}}]{fishbacher23}
{Fischbacher}, S., {Kacprzak}, T., {Blazek}, J., \& {Refregier}, A. 2023,
  \href{http://dx.doi.org/10.1088/1475-7516/2023/01/033}{\color{magenta}\jcap},
  \href{https://ui.adsabs.harvard.edu/abs/2023JCAP...01..033F}{2023, 033}

\bibitem[{{Gao} {et~al.}(2008){Gao}, {Navarro}, {Cole}, {Frenk}, {White},
  {Springel}, {Jenkins}, \& {Neto}}]{gao08}
{Gao}, L., {Navarro}, J.~F., {Cole}, S., {et~al.} 2008,
  \href{http://dx.doi.org/10.1111/j.1365-2966.2008.13277.x}{\color{magenta}\mnras},
  \href{http://adsabs.harvard.edu/abs/2008MNRAS.387..536G}{387, 536}

\bibitem[{{Garc{\'\i}a} {et~al.}(2023){Garc{\'\i}a}, {Salazar}, {Rozo},
  {Adhikari}, {Aung}, {Diemer}, {Nagai}, \& {Wolfe}}]{garcia23}
{Garc{\'\i}a}, R., {Salazar}, E., {Rozo}, E., {et~al.} 2023,
  \href{http://dx.doi.org/10.1093/mnras/stad660}{\color{magenta}\mnras},
  \href{https://ui.adsabs.harvard.edu/abs/2023MNRAS.521.2464G}{521, 2464}

\bibitem[{{Giocoli} {et~al.}(2010){Giocoli}, {Bartelmann}, {Sheth}, \&
  {Cacciato}}]{giocoli10b}
{Giocoli}, C., {Bartelmann}, M., {Sheth}, R.~K., \& {Cacciato}, M. 2010,
  \href{http://dx.doi.org/10.1111/j.1365-2966.2010.17108.x}{\color{magenta}\mnras},
  \href{http://adsabs.harvard.edu/abs/2010MNRAS.408..300G}{408, 300}

\bibitem[{{Giocoli} {et~al.}(2021){Giocoli}, {Marulli}, {Moscardini}, {Sereno},
  {Veropalumbo}, {Gigante}, \& {et al.}}]{giocoli21}
{Giocoli}, C., {Marulli}, F., {Moscardini}, L., {et~al.} 2021,
  \href{http://dx.doi.org/10.1051/0004-6361/202140795}{\color{magenta}\aap},
  \href{https://ui.adsabs.harvard.edu/abs/2021A&A...653A..19G}{653, A19}

\bibitem[{{Giocoli} {et~al.}(2012a){Giocoli}, {Meneghetti}, {Bartelmann},
  {Moscardini}, \& {Boldrin}}]{giocoli12a}
{Giocoli}, C., {Meneghetti}, M., {Bartelmann}, M., {Moscardini}, L., \&
  {Boldrin}, M. 2012a,
  \href{http://dx.doi.org/10.1111/j.1365-2966.2012.20558.x}{\color{magenta}\mnras},
  \href{http://adsabs.harvard.edu/abs/2012MNRAS.421.3343G}{421, 3343}

\bibitem[{{Giocoli} {et~al.}(2007){Giocoli}, {Moreno}, {Sheth}, \&
  {Tormen}}]{giocoli07}
{Giocoli}, C., {Moreno}, J., {Sheth}, R.~K., \& {Tormen}, G. 2007,
  \href{http://dx.doi.org/10.1111/j.1365-2966.2007.11520.x}{\color{magenta}\mnras},
  \href{http://adsabs.harvard.edu/abs/2007MNRAS.376..977G}{376, 977}

\bibitem[{{Giocoli} {et~al.}(2012b){Giocoli}, {Tormen}, \&
  {Sheth}}]{giocoli12b}
{Giocoli}, C., {Tormen}, G., \& {Sheth}, R.~K. 2012b,
  \href{http://dx.doi.org/10.1111/j.1365-2966.2012.20594.x}{\color{magenta}\mnras},
  \href{http://adsabs.harvard.edu/abs/2012MNRAS.422..185G}{422, 185}

\bibitem[{{Giocoli} {et~al.}(2008){Giocoli}, {Tormen}, \& {van den
  Bosch}}]{giocoli08b}
{Giocoli}, C., {Tormen}, G., \& {van den Bosch}, F.~C. 2008,
  \href{http://dx.doi.org/10.1111/j.1365-2966.2008.13182.x}{\color{magenta}\mnras},
  \href{http://adsabs.harvard.edu/abs/2008MNRAS.386.2135G}{386, 2135}

\bibitem[{{Gunn} \& {Gott}(1972)}]{gunn72}
{Gunn}, J.~E. \& {Gott}, J.~Richard, I. 1972,
  \href{http://dx.doi.org/10.1086/151605}{\color{magenta}\apj},
  \href{https://ui.adsabs.harvard.edu/abs/1972ApJ...176....1G}{176, 1}

\bibitem[{{Heymans} {et~al.}(2012{\natexlab{a}}){Heymans}, {Rowe}, {Hoekstra},
  {Miller}, {Erben}, {Kitching}, \& {van Waerbeke}}]{heymans12}
{Heymans}, C., {Rowe}, B., {Hoekstra}, H., {et~al.} 2012{\natexlab{a}},
  \href{http://dx.doi.org/10.1111/j.1365-2966.2011.20312.x}{\color{magenta}\mnras},
  \href{http://adsabs.harvard.edu/abs/2012MNRAS.421..381H}{421, 381}

\bibitem[{{Heymans} {et~al.}(2012{\natexlab{b}}){Heymans}, {Van Waerbeke},
  {Miller}, {Erben}, {Hildebrandt}, {Hoekstra}, {Kitching}, \& {et
  al.}}]{heymans12b}
{Heymans}, C., {Van Waerbeke}, L., {Miller}, L., {et~al.} 2012{\natexlab{b}},
  \href{http://dx.doi.org/10.1111/j.1365-2966.2012.21952.x}{\color{magenta}\mnras},
  \href{https://ui.adsabs.harvard.edu/abs/2012MNRAS.427..146H}{427, 146}

\bibitem[{{Hildebrandt} {et~al.}(2012){Hildebrandt}, {Erben}, {Kuijken}, {van
  Waerbeke}, {Heymans}, {Coupon}, {Benjamin}, \& {et al.}}]{hildebrandt12}
{Hildebrandt}, H., {Erben}, T., {Kuijken}, K., {et~al.} 2012,
  \href{http://dx.doi.org/10.1111/j.1365-2966.2012.20468.x}{\color{magenta}\mnras},
  \href{http://adsabs.harvard.edu/abs/2012MNRAS.421.2355H}{421, 2355}

\bibitem[{{Hildebrandt} {et~al.}(2017){Hildebrandt}, {Viola}, {Heymans},
  {Joudaki}, {Kuijken}, {Blake}, {Erben}, {Joachimi}, \& {et
  al.}}]{hildebrandt17}
{Hildebrandt}, H., {Viola}, M., {Heymans}, C., {et~al.} 2017,
  \href{http://dx.doi.org/10.1093/mnras/stw2805}{\color{magenta}\mnras},
  \href{http://adsabs.harvard.edu/abs/2017MNRAS.465.1454H}{465, 1454}

\bibitem[{{Ingoglia} {et~al.}(2022){Ingoglia}, {Covone}, {Sereno}, {Giocoli},
  {Bardelli}, {Bellagamba}, \& {et al.}}]{ingoglia22}
{Ingoglia}, L., {Covone}, G., {Sereno}, M., {et~al.} 2022,
  \href{http://dx.doi.org/10.1093/mnras/stac046}{\color{magenta}\mnras},
  \href{https://ui.adsabs.harvard.edu/abs/2022MNRAS.511.1484I}{511, 1484}

\bibitem[{{Johnston} {et~al.}(2007){Johnston}, {Sheldon}, {Wechsler}, {Rozo},
  {Koester}, {Frieman}, {McKay}, {Evrard}, {Becker}, \& {Annis}}]{johnston07}
{Johnston}, D.~E., {Sheldon}, E.~S., {Wechsler}, R.~H., {et~al.} 2007,
  \href{https://ui.adsabs.harvard.edu/abs/2007arXiv0709.1159J}{arXiv e-prints,
  arXiv:0709.1159}

\bibitem[{{Kuijken}(2011)}]{omegacameso11}
{Kuijken}, K. 2011, The Messenger,
  \href{https://ui.adsabs.harvard.edu/abs/2011Msngr.146....8K}{146, 8}

\bibitem[{{Kuijken} {et~al.}(2015){Kuijken}, {Heymans}, {Hildebrandt},
  {Nakajima}, {Erben}, {de Jong}, {Viola}, {Choi}, \& {et al.}}]{kuijken15}
{Kuijken}, K., {Heymans}, C., {Hildebrandt}, H., {et~al.} 2015,
  \href{http://dx.doi.org/10.1093/mnras/stv2140}{\color{magenta}\mnras},
  \href{https://ui.adsabs.harvard.edu/abs/2015MNRAS.454.3500K}{454, 3500}

\bibitem[{{Lacey} \& {Cole}(1993)}]{lacey93}
{Lacey}, C. \& {Cole}, S. 1993, \mnras,
  \href{http://adsabs.harvard.edu/abs/1993MNRAS.262..627L}{262, 627}

\bibitem[{{Laureijs} {et~al.}(2011){Laureijs}, {Amiaux}, {Arduini},
  {Augu{\`e}res}, {Brinchmann}, {Cole}, {Cropper}, {Dabin}, {Duvet}, \&
  et~al.}]{euclidredbook}
{Laureijs}, R., {Amiaux}, J., {Arduini}, S., {et~al.} 2011,
  \href{http://adsabs.harvard.edu/abs/2011arXiv1110.3193L}{arXiv:1110.3193}

\bibitem[{{Lesci} {et~al.}(2022{\natexlab{a}}){Lesci}, {Marulli}, {Moscardini},
  {Sereno}, {Veropalumbo}, {Maturi}, {Giocoli}, {Radovich}, {Bellagamba},
  {Roncarelli}, {Bardelli}, {Contarini}, {Covone}, {Ingoglia}, {Nanni}, \&
  {Puddu}}]{lesci22}
{Lesci}, G.~F., {Marulli}, F., {Moscardini}, L., {et~al.} 2022{\natexlab{a}},
  \href{http://dx.doi.org/10.1051/0004-6361/202040194}{\color{magenta}\aap},
  \href{https://ui.adsabs.harvard.edu/abs/2022A&A...659A..88L}{659, A88}

\bibitem[{{Lesci} {et~al.}(2022{\natexlab{b}}){Lesci}, {Nanni}, {Marulli},
  {Moscardini}, {Veropalumbo}, {Maturi}, {Sereno}, {Radovich}, {Bellagamba},
  {Roncarelli}, {Bardelli}, {Castignani}, {Covone}, {Giocoli}, {Ingoglia}, \&
  {Puddu}}]{lesci22b}
{Lesci}, G.~F., {Nanni}, L., {Marulli}, F., {et~al.} 2022{\natexlab{b}},
  \href{http://dx.doi.org/10.1051/0004-6361/202243538}{\color{magenta}\aap},
  \href{https://ui.adsabs.harvard.edu/abs/2022A&A...665A.100L}{665, A100}

\bibitem[{Lewis {et~al.}(2000)Lewis, Challinor, \& Lasenby}]{camb}
Lewis, A., Challinor, A., \& Lasenby, A. 2000,
  \href{http://dx.doi.org/10.1086/309179}{\color{magenta}Astrophys. J.}, 538,
  538

\bibitem[{{Liske} {et~al.}(2015){Liske}, {Baldry}, {Driver}, {Tuffs},
  {Alpaslan}, {Andrae}, {Brough}, {Cluver}, \& {etl al.}}]{liske15}
{Liske}, J., {Baldry}, I.~K., {Driver}, S.~P., {et~al.} 2015,
  \href{http://dx.doi.org/10.1093/mnras/stv1436}{\color{magenta}\mnras},
  \href{https://ui.adsabs.harvard.edu/abs/2015MNRAS.452.2087L}{452, 2087}

\bibitem[{{Macci{\`o}} {et~al.}(2008){Macci{\`o}}, {Dutton}, \& {van den
  Bosch}}]{maccio08}
{Macci{\`o}}, A.~V., {Dutton}, A.~A., \& {van den Bosch}, F.~C. 2008,
  \href{http://dx.doi.org/10.1111/j.1365-2966.2008.14029.x}{\color{magenta}\mnras},
  \href{http://adsabs.harvard.edu/abs/2008MNRAS.391.1940M}{391, 1940}

\bibitem[{{Macci{\`o}} {et~al.}(2007){Macci{\`o}}, {Dutton}, {van den Bosch},
  {Moore}, {Potter}, \& {Stadel}}]{maccio07}
{Macci{\`o}}, A.~V., {Dutton}, A.~A., {van den Bosch}, F.~C., {et~al.} 2007,
  \href{http://dx.doi.org/10.1111/j.1365-2966.2007.11720.x}{\color{magenta}\mnras},
  \href{http://adsabs.harvard.edu/abs/2007MNRAS.378...55M}{378, 55}

\bibitem[{{Marulli} {et~al.}(2016){Marulli}, {Veropalumbo}, \&
  {Moresco}}]{marulli16}
{Marulli}, F., {Veropalumbo}, A., \& {Moresco}, M. 2016,
  \href{http://dx.doi.org/10.1016/j.ascom.2016.01.005}{\color{magenta}Astronomy
  and Computing},
  \href{https://ui.adsabs.harvard.edu/abs/2016A&C....14...35M}{14, 35}

\bibitem[{{Maturi} {et~al.}(2019){Maturi}, {Bellagamba}, {Radovich},
  {Roncarelli}, {Sereno}, {Moscardini}, {Bardelli}, \& {Puddu}}]{maturi19}
{Maturi}, M., {Bellagamba}, F., {Radovich}, M., {et~al.} 2019,
  \href{http://dx.doi.org/10.1093/mnras/stz294}{\color{magenta}\mnras},
  \href{https://ui.adsabs.harvard.edu/abs/2019MNRAS.485..498M}{485, 498}

\bibitem[{{McClintock} {et~al.}(2019){McClintock}, {Varga}, {Gruen}, {Rozo},
  {Rykoff}, {Shin}, {Melchior}, {DeRose}, {Seitz}, {Dietrich}, {Sheldon},
  {Zhang}, {von der Linden}, {Jeltema}, {Mantz}, {Romer}, {Allen}, {Becker},
  {Bermeo}, {Bhargava}, {Costanzi}, {Everett}, {Farahi}, {Hamaus}, {Hartley},
  {Hollowood}, {Hoyle}, {Israel}, {Li}, {MacCrann}, {Morris}, {Palmese},
  {Plazas}, {Pollina}, {Rau}, {Simet}, {Soares-Santos}, {Troxel}, {Vergara
  Cervantes}, {Wechsler}, {Zuntz}, {Abbott}, {Abdalla}, {Allam}, {Annis},
  {Avila}, {Bridle}, {Brooks}, {Burke}, {Carnero Rosell}, {Carrasco Kind},
  {Carretero}, {Castander}, {Crocce}, {Cunha}, {D'Andrea}, {da Costa}, {Davis},
  {De Vicente}, {Diehl}, {Doel}, {Drlica-Wagner}, {Evrard}, {Flaugher},
  {Fosalba}, {Frieman}, {Garc{\'\i}a-Bellido}, {Gaztanaga}, {Gerdes},
  {Giannantonio}, {Gruendl}, {Gutierrez}, {Honscheid}, {James}, {Kirk},
  {Krause}, {Kuehn}, {Lahav}, {Li}, {Lima}, {March}, {Marshall}, {Menanteau},
  {Miquel}, {Mohr}, {Nord}, {Ogando}, {Roodman}, {Sanchez}, {Scarpine},
  {Schindler}, {Sevilla-Noarbe}, {Smith}, {Smith}, {Sobreira}, {Suchyta},
  {Swanson}, {Tarle}, {Tucker}, {Vikram}, {Walker}, {Weller}, \& {DES
  Collaboration}}]{mcclintock19}
{McClintock}, T., {Varga}, T.~N., {Gruen}, D., {et~al.} 2019,
  \href{http://dx.doi.org/10.1093/mnras/sty2711}{\color{magenta}\mnras},
  \href{https://ui.adsabs.harvard.edu/abs/2019MNRAS.482.1352M}{482, 1352}

\bibitem[{{Miller} {et~al.}(2013){Miller}, {Heymans}, {Kitching}, {van
  Waerbeke}, {Erben}, {Hildebrandt}, {Hoekstra}, \& {et al.}}]{miller13}
{Miller}, L., {Heymans}, C., {Kitching}, T.~D., {et~al.} 2013,
  \href{http://dx.doi.org/10.1093/mnras/sts454}{\color{magenta}\mnras},
  \href{http://adsabs.harvard.edu/abs/2013MNRAS.429.2858M}{429, 2858}

\bibitem[{{Miller} {et~al.}(2007){Miller}, {Kitching}, {Heymans}, {Heavens}, \&
  {van Waerbeke}}]{miller07}
{Miller}, L., {Kitching}, T.~D., {Heymans}, C., {Heavens}, A.~F., \& {van
  Waerbeke}, L. 2007,
  \href{http://dx.doi.org/10.1111/j.1365-2966.2007.12363.x}{\color{magenta}\mnras},
  \href{https://ui.adsabs.harvard.edu/abs/2007MNRAS.382..315M}{382, 315}

\bibitem[{{More} {et~al.}(2015){More}, {Diemer}, \& {Kravtsov}}]{more15}
{More}, S., {Diemer}, B., \& {Kravtsov}, A.~V. 2015,
  \href{http://dx.doi.org/10.1088/0004-637X/810/1/36}{\color{magenta}\apj},
  \href{https://ui.adsabs.harvard.edu/abs/2015ApJ...810...36M}{810, 36}

\bibitem[{{More} {et~al.}(2016){More}, {Miyatake}, {Takada}, {Diemer},
  {Kravtsov}, {Dalal}, {More}, {Murata}, {Mandelbaum}, {Rozo}, {Rykoff},
  {Oguri}, \& {Spergel}}]{more16}
{More}, S., {Miyatake}, H., {Takada}, M., {et~al.} 2016,
  \href{http://dx.doi.org/10.3847/0004-637X/825/1/39}{\color{magenta}\apj},
  \href{https://ui.adsabs.harvard.edu/abs/2016ApJ...825...39M}{825, 39}

\bibitem[{{Murata} {et~al.}(2020){Murata}, {Sunayama}, {Oguri}, {More},
  {Nishizawa}, {Nishimichi}, \& {Osato}}]{murata20}
{Murata}, R., {Sunayama}, T., {Oguri}, M., {et~al.} 2020,
  \href{http://dx.doi.org/10.1093/pasj/psaa041}{\color{magenta}\pasj},
  \href{https://ui.adsabs.harvard.edu/abs/2020PASJ...72...64M}{72, 64}

\bibitem[{{Navarro} {et~al.}(1996){Navarro}, {Frenk}, \& {White}}]{navarro96}
{Navarro}, J.~F., {Frenk}, C.~S., \& {White}, S.~D.~M. 1996,
  \href{http://dx.doi.org/10.1086/177173}{\color{magenta}\apj},
  \href{http://adsabs.harvard.edu/abs/1996ApJ...462..563N}{462, 563}

\bibitem[{{Navarro} {et~al.}(1997){Navarro}, {Frenk}, \& {White}}]{navarro97}
{Navarro}, J.~F., {Frenk}, C.~S., \& {White}, S.~D.~M. 1997,
  \href{http://dx.doi.org/10.1086/304888}{\color{magenta}\apj},
  \href{http://adsabs.harvard.edu/abs/1997ApJ...490..493N}{490, 493}

\bibitem[{{Nelson} {et~al.}(2018){Nelson}, {Pillepich}, {Springel},
  {Weinberger}, {Hernquist}, {Pakmor}, {Genel}, {Torrey}, {Vogelsberger},
  {Kauffmann}, {Marinacci}, \& {Naiman}}]{nelson18}
{Nelson}, D., {Pillepich}, A., {Springel}, V., {et~al.} 2018,
  \href{http://dx.doi.org/10.1093/mnras/stx3040}{\color{magenta}\mnras},
  \href{https://ui.adsabs.harvard.edu/abs/2018MNRAS.475..624N}{475, 624}

\bibitem[{{Neto} {et~al.}(2007){Neto}, {Gao}, {Bett}, {Cole}, {Navarro},
  {Frenk}, {White}, {Springel}, \& {Jenkins}}]{neto07}
{Neto}, A.~F., {Gao}, L., {Bett}, P., {et~al.} 2007,
  \href{http://dx.doi.org/10.1111/j.1365-2966.2007.12381.x}{\color{magenta}\mnras},
  \href{http://adsabs.harvard.edu/abs/2007MNRAS.381.1450N}{381, 1450}

\bibitem[{{Oguri} \& {Hamana}(2011)}]{oguri11b}
{Oguri}, M. \& {Hamana}, T. 2011,
  \href{http://dx.doi.org/10.1111/j.1365-2966.2011.18481.x}{\color{magenta}\mnras},
  \href{https://ui.adsabs.harvard.edu/abs/2011MNRAS.414.1851O}{414, 1851}

\bibitem[{{Oguri} \& {Takada}(2011)}]{oguri11}
{Oguri}, M. \& {Takada}, M. 2011,
  \href{http://dx.doi.org/10.1103/PhysRevD.83.023008}{\color{magenta}\prd},
  \href{http://adsabs.harvard.edu/abs/2011PhRvD..83b3008O}{83, 023008}

\bibitem[{{Pillepich} {et~al.}(2018){Pillepich}, {Nelson}, {Hernquist},
  {Springel}, {Pakmor}, {Torrey}, {Weinberger}, {Genel}, {Naiman}, {Marinacci},
  \& {Vogelsberger}}]{pillepich18}
{Pillepich}, A., {Nelson}, D., {Hernquist}, L., {et~al.} 2018,
  \href{http://dx.doi.org/10.1093/mnras/stx3112}{\color{magenta}\mnras},
  \href{https://ui.adsabs.harvard.edu/abs/2018MNRAS.475..648P}{475, 648}

\bibitem[{{Pizzardo} {et~al.}(2024){Pizzardo}, {Geller}, {Kenyon}, \&
  {Damjanov}}]{pizzardo24}
{Pizzardo}, M., {Geller}, M.~J., {Kenyon}, S.~J., \& {Damjanov}, I. 2024,
  \href{http://dx.doi.org/10.1051/0004-6361/202348643}{\color{magenta}\aap},
  \href{https://ui.adsabs.harvard.edu/abs/2024A&A...683A..82P}{683, A82}

\bibitem[{{Planck Collaboration} {et~al.}(2020){Planck Collaboration},
  {Aghanim}, {Akrami}, {Ashdown}, {Aumont}, {Baccigalupi}, {Ballardini}, \& {et
  al.}}]{planck18}
{Planck Collaboration}, {Aghanim}, N., {Akrami}, Y., {et~al.} 2020,
  \href{http://dx.doi.org/10.1051/0004-6361/201833910}{\color{magenta}\aap},
  \href{https://ui.adsabs.harvard.edu/abs/2020A&A...641A...6P}{641, A6}

\bibitem[{{Press} \& {Schechter}(1974)}]{press74}
{Press}, W.~H. \& {Schechter}, P. 1974, \apj,
  \href{http://adsabs.harvard.edu/abs/1974ApJ...187..425P}{187, 425}

\bibitem[{{Puddu} {et~al.}(2021){Puddu}, {Radovich}, {Sereno}, {Bardelli},
  {Maturi}, {Moscardini}, {Bellagamba}, {Giocoli}, {Marulli}, \&
  {Roncarelli}}]{puddu21}
{Puddu}, E., {Radovich}, M., {Sereno}, M., {et~al.} 2021,
  \href{http://dx.doi.org/10.1051/0004-6361/202039259}{\color{magenta}\aap},
  \href{https://ui.adsabs.harvard.edu/abs/2021A&A...645A...9P}{645, A9}

\bibitem[{{Radovich} {et~al.}(2020){Radovich}, {Tortora}, {Bellagamba},
  {Maturi}, {Moscardini}, {Puddu}, {Roncarelli}, {Roy}, {Bardelli}, {Marulli},
  {Sereno}, {Getman}, \& {Napolitano}}]{radovich20}
{Radovich}, M., {Tortora}, C., {Bellagamba}, F., {et~al.} 2020,
  \href{http://dx.doi.org/10.1093/mnras/staa2705}{\color{magenta}\mnras},
  \href{https://ui.adsabs.harvard.edu/abs/2020MNRAS.498.4303R}{498, 4303}

\bibitem[{{Rana} {et~al.}(2023){Rana}, {More}, {Miyatake}, {Grandis}, {Klein},
  {Bulbul}, {Chiu}, {Miyazaki}, \& {Bahcall}}]{rana23}
{Rana}, D., {More}, S., {Miyatake}, H., {et~al.} 2023,
  \href{http://dx.doi.org/10.1093/mnras/stad1239}{\color{magenta}\mnras},
  \href{https://ui.adsabs.harvard.edu/abs/2023MNRAS.522.4181R}{522, 4181}

\bibitem[{{Romanello} {et~al.}(2024){Romanello}, {Marulli}, {Moscardini},
  {Lesci}, {Sartoris}, {Contarini}, {Giocoli}, {Bardelli}, {Busillo},
  {Castignani}, {Covone}, {Ingoglia}, {Maturi}, {Puddu}, {Radovich},
  {Roncarelli}, \& {Sereno}}]{romanello24}
{Romanello}, M., {Marulli}, F., {Moscardini}, L., {et~al.} 2024,
  \href{http://dx.doi.org/10.1051/0004-6361/202348305}{\color{magenta}\aap},
  \href{https://ui.adsabs.harvard.edu/abs/2024A&A...682A..72R}{682, A72}

\bibitem[{{Sartoris} {et~al.}(2016){Sartoris}, {Biviano}, {Fedeli}, {Bartlett},
  {Borgani}, {Costanzi}, {Giocoli}, {Moscardini}, {Weller}, {Ascaso},
  {Bardelli}, {Maurogordato}, \& {Viana}}]{sartoris16}
{Sartoris}, B., {Biviano}, A., {Fedeli}, C., {et~al.} 2016,
  \href{http://dx.doi.org/10.1093/mnras/stw630}{\color{magenta}\mnras},
  \href{http://adsabs.harvard.edu/abs/2016MNRAS.459.1764S}{459, 1764}

\bibitem[{{Schaye} {et~al.}(2023){Schaye}, {Kugel}, {Schaller}, {Helly},
  {Braspenning}, {Elbers}, {McCarthy}, {van Daalen}, {Vandenbroucke}, {Frenk},
  {Kwan}, {Salcido}, {Bah{\'e}}, {Borrow}, {Chaikin}, {Hahn}, {Hu{\v{s}}ko},
  {Jenkins}, {Lacey}, \& {Nobels}}]{shaye23}
{Schaye}, J., {Kugel}, R., {Schaller}, M., {et~al.} 2023,
  \href{http://dx.doi.org/10.1093/mnras/stad2419}{\color{magenta}\mnras},
  \href{https://ui.adsabs.harvard.edu/abs/2023MNRAS.526.4978S}{526, 4978}

\bibitem[{{Schechter}(1976)}]{schechter76}
{Schechter}, P. 1976,
  \href{http://dx.doi.org/10.1086/154079}{\color{magenta}\apj},
  \href{https://ui.adsabs.harvard.edu/abs/1976ApJ...203..297S}{203, 297}

\bibitem[{{Sereno} {et~al.}(2017){Sereno}, {Covone}, {Izzo}, {Ettori},
  {Coupon}, \& {Lieu}}]{sereno17}
{Sereno}, M., {Covone}, G., {Izzo}, L., {et~al.} 2017,
  \href{http://dx.doi.org/10.1093/mnras/stx2085}{\color{magenta}\mnras},
  \href{https://ui.adsabs.harvard.edu/abs/2017MNRAS.472.1946S}{472, 1946}

\bibitem[{{Sereno} \& {Ettori}(2015)}]{sereno15c}
{Sereno}, M. \& {Ettori}, S. 2015,
  \href{http://dx.doi.org/10.1093/mnras/stv814}{\color{magenta}\mnras},
  \href{https://ui.adsabs.harvard.edu/abs/2015MNRAS.450.3675S}{450, 3675}

\bibitem[{{Sereno} {et~al.}(2018){Sereno}, {Giocoli}, {Izzo}, {Marulli},
  {Veropalumbo}, {Ettori}, {Moscardini}, {Covone}, {Ferragamo}, {Barrena}, \&
  {Streblyanska}}]{sereno18a}
{Sereno}, M., {Giocoli}, C., {Izzo}, L., {et~al.} 2018,
  \href{http://dx.doi.org/10.1038/s41550-018-0508-y}{\color{magenta}Nature
  Astronomy}, \href{https://ui.adsabs.harvard.edu/abs/2018NatAs...2..744S}{2,
  744}

\bibitem[{{Sheth} {et~al.}(2001){Sheth}, {Mo}, \& {Tormen}}]{sheth01b}
{Sheth}, R.~K., {Mo}, H.~J., \& {Tormen}, G. 2001,
  \href{http://dx.doi.org/10.1046/j.1365-8711.2001.04006.x}{\color{magenta}\mnras},
  \href{http://adsabs.harvard.edu/abs/2001MNRAS.323....1S}{323, 1}

\bibitem[{{Sheth} \& {Tormen}(1999)}]{sheth99b}
{Sheth}, R.~K. \& {Tormen}, G. 1999, \mnras,
  \href{http://adsabs.harvard.edu/abs/1999MNRAS.308..119S}{308, 119}

\bibitem[{{Shin} {et~al.}(2019){Shin}, {Adhikari}, {Baxter}, {Chang}, {Jain},
  {Battaglia}, {Bleem}, {Bocquet}, {DeRose}, {Gruen}, {Hilton}, {Kravtsov},
  {McClintock}, {Rozo}, {Rykoff}, {Varga}, {Wechsler}, {Wu}, {Zhang}, {Aiola},
  {Allam}, {Bechtol}, {Benson}, {Bertin}, {Bond}, {Brodwin}, {Brooks},
  {Buckley-Geer}, {Burke}, {Carlstrom}, {Carnero Rosell}, {Carrasco Kind},
  {Carretero}, {Castander}, {Choi}, {Cunha}, {Crawford}, {da Costa}, {De
  Vicente}, {Desai}, {Devlin}, {Dietrich}, {Doel}, {Dunkley}, {Eifler},
  {Evrard}, {Flaugher}, {Fosalba}, {Gallardo}, {Garc{\'\i}a-Bellido},
  {Gaztanaga}, {Gerdes}, {Gralla}, {Gruendl}, {Gschwend}, {Gupta}, {Gutierrez},
  {Hartley}, {Hill}, {Ho}, {Hollowood}, {Honscheid}, {Hoyle}, {Huffenberger},
  {Hughes}, {James}, {Jeltema}, {Kim}, {Krause}, {Kuehn}, {Lahav}, {Lima},
  {Madhavacheril}, {Maia}, {Marshall}, {Maurin}, {McMahon}, {Menanteau},
  {Miller}, {Miquel}, {Mohr}, {Naess}, {Nati}, {Newburgh}, {Niemack}, {Ogando},
  {Page}, {Partridge}, {Patil}, {Plazas}, {Rapetti}, {Reichardt}, {Romer},
  {Sanchez}, {Scarpine}, {Schindler}, {Serrano}, {Smith}, {Smith},
  {Soares-Santos}, {Sobreira}, {Staggs}, {Stark}, {Stein}, {Suchyta},
  {Swanson}, {Tarle}, {Thomas}, {van Engelen}, {Wollack}, \& {Xu}}]{shin19}
{Shin}, T., {Adhikari}, S., {Baxter}, E.~J., {et~al.} 2019,
  \href{http://dx.doi.org/10.1093/mnras/stz1434}{\color{magenta}\mnras},
  \href{https://ui.adsabs.harvard.edu/abs/2019MNRAS.487.2900S}{487, 2900}

\bibitem[{{Shin} {et~al.}(2021){Shin}, {Jain}, {Adhikari}, {Baxter}, {Chang},
  {Pandey}, {Salcedo}, {Weinberg}, {Amsellem}, {Battaglia}, {Belyakov},
  {Dacunha}, {Goldstein}, {Kravtsov}, {Varga}, {Abbott}, {Aguena}, {Alarcon},
  {Allam}, {Amon}, {Andrade-Oliveira}, {Annis}, {Bacon}, {Bechtol}, {Becker},
  {Bernstein}, {Bertin}, {Bocquet}, {Bond}, {Brooks}, {Buckley-Geer}, {Burke},
  {Campos}, {Rosell}, {Kind}, {Carretero}, {Chen}, {Choi}, {Costanzi}, {da
  Costa}, {DeRose}, {Desai}, {De Vicente}, {Devlin}, {Diehl}, {Dietrich},
  {Dodelson}, {Doel}, {Doux}, {Drlica-Wagner}, {Eckert}, {Elvin-Poole},
  {Everett}, {Ferraro}, {Ferrero}, {Fert{\'e}}, {Flaugher}, {Frieman},
  {Gallardo}, {Gatti}, {Gaztanaga}, {Gerdes}, {Gruen}, {Gruendl}, {Gutierrez},
  {Harrison}, {Hartley}, {Hill}, {Hilton}, {Hinton}, {Hollowood}, {Hughes},
  {James}, {Jarvis}, {Jeltema}, {Koopman}, {Krause}, {Kuehn}, {Kuropatkin},
  {Lahav}, {Lima}, {Lokken}, {MacCrann}, {Madhavacheril}, {Maia}, {McCullough},
  {McMahon}, {Melchior}, {Menanteau}, {Miquel}, {Mohr}, {Moodley}, {Morgan},
  {Myles}, {Nati}, {Navarro-Alsina}, {Niemack}, {Ogando}, {Page}, {Palmese},
  {Partridge}, {Paz-Chinch{\'o}n}, {Pereira}, {Pieres}, {Malag{\'o}n}, {Prat},
  {Raveri}, {Rodriguez-Monroy}, {Rollins}, {Romer}, {Rykoff}, {Salatino},
  {S{\'a}nchez}, {Sanchez}, {Santiago}, {Scarpine}, {Schillaci}, {Secco},
  {Serrano}, {Sevilla-Noarbe}, {Sheldon}, {Sherwin}, {Sif{\'o}n}, {Smith},
  {Soares-Santos}, {Staggs}, {Suchyta}, {Swanson}, {Tarle}, {Thomas}, {To},
  {Troxel}, {Tutusaus}, {Vavagiakis}, {Weller}, {Wollack}, {Yanny}, {Yin}, \&
  {Zhang}}]{shin21}
{Shin}, T., {Jain}, B., {Adhikari}, S., {et~al.} 2021,
  \href{http://dx.doi.org/10.1093/mnras/stab2505}{\color{magenta}\mnras},
  \href{https://ui.adsabs.harvard.edu/abs/2021MNRAS.507.5758S}{507, 5758}

\bibitem[{{Shin} \& {Diemer}(2023)}]{shin23}
{Shin}, T.-h. \& {Diemer}, B. 2023,
  \href{http://dx.doi.org/10.1093/mnras/stad860}{\color{magenta}\mnras},
  \href{https://ui.adsabs.harvard.edu/abs/2023MNRAS.521.5570S}{521, 5570}

\bibitem[{{Springel}(2010)}]{springel10}
{Springel}, V. 2010,
  \href{http://dx.doi.org/10.1111/j.1365-2966.2009.15715.x}{\color{magenta}\mnras},
  \href{http://adsabs.harvard.edu/abs/2010MNRAS.401..791S}{401, 791}

\bibitem[{{Springel} {et~al.}(2018){Springel}, {Pakmor}, {Pillepich},
  {Weinberger}, {Nelson}, {Hernquist}, {Vogelsberger}, {Genel}, {Torrey},
  {Marinacci}, \& {Naiman}}]{springel18}
{Springel}, V., {Pakmor}, R., {Pillepich}, A., {et~al.} 2018,
  \href{http://dx.doi.org/10.1093/mnras/stx3304}{\color{magenta}\mnras},
  \href{https://ui.adsabs.harvard.edu/abs/2018MNRAS.475..676S}{475, 676}

\bibitem[{{Springel} {et~al.}(2001){Springel}, {White}, {Tormen}, \&
  {Kauffmann}}]{springel01b}
{Springel}, V., {White}, S.~D.~M., {Tormen}, G., \& {Kauffmann}, G. 2001,
  \href{http://dx.doi.org/10.1046/j.1365-8711.2001.04912.x}{\color{magenta}\mnras},
  \href{http://adsabs.harvard.edu/abs/2001MNRAS.328..726S}{328, 726}

\bibitem[{{Tinker} {et~al.}(2008){Tinker}, {Kravtsov}, {Klypin}, {Abazajian},
  {Warren}, {Yepes}, {Gottl{\"o}ber}, \& {Holz}}]{tinker08}
{Tinker}, J., {Kravtsov}, A.~V., {Klypin}, A., {et~al.} 2008,
  \href{http://dx.doi.org/10.1086/591439}{\color{magenta}\apj},
  \href{http://adsabs.harvard.edu/abs/2008ApJ...688..709T}{688, 709}

\bibitem[{{Tinker} {et~al.}(2010){Tinker}, {Robertson}, {Kravtsov}, {Klypin},
  {Warren}, {Yepes}, \& {Gottl{\"o}ber}}]{tinker10}
{Tinker}, J.~L., {Robertson}, B.~E., {Kravtsov}, A.~V., {et~al.} 2010,
  \href{http://dx.doi.org/10.1088/0004-637X/724/2/878}{\color{magenta}\apj},
  \href{http://adsabs.harvard.edu/abs/2010ApJ...724..878T}{724, 878}

\bibitem[{{Tormen}(1998)}]{tormen98a}
{Tormen}, G. 1998, \mnras,
  \href{http://adsabs.harvard.edu/abs/1998MNRAS.297..648T}{297, 648}

\bibitem[{{Towler} {et~al.}(2024){Towler}, {Kay}, {Schaye}, {Kugel},
  {Schaller}, {Braspenning}, {Elbers}, {Frenk}, {Kwan}, {Salcido}, {van
  Daalen}, {Vandenbroucke}, \& {Altamura}}]{towler24}
{Towler}, I., {Kay}, S.~T., {Schaye}, J., {et~al.} 2024,
  \href{http://dx.doi.org/10.1093/mnras/stae654}{\color{magenta}\mnras},
  \href{https://ui.adsabs.harvard.edu/abs/2024MNRAS.529.2017T}{529, 2017}

\bibitem[{{Umetsu} \& {Diemer}(2017)}]{umetsu17}
{Umetsu}, K. \& {Diemer}, B. 2017,
  \href{http://dx.doi.org/10.3847/1538-4357/aa5c90}{\color{magenta}\apj},
  \href{https://ui.adsabs.harvard.edu/abs/2017ApJ...836..231U}{836, 231}

\bibitem[{{van den Bosch}(2002)}]{vandenbosch02}
{van den Bosch}, F.~C. 2002,
  \href{http://dx.doi.org/10.1046/j.1365-8711.2002.05171.x}{\color{magenta}\mnras},
  \href{http://adsabs.harvard.edu/abs/2002MNRAS.331...98V}{331, 98}

\bibitem[{{Viola} {et~al.}(2015){Viola}, {Cacciato}, {Brouwer}, {Kuijken},
  {Hoekstra}, {Norberg}, {Robotham}, \& {etl al.}}]{viola15}
{Viola}, M., {Cacciato}, M., {Brouwer}, M., {et~al.} 2015,
  \href{http://dx.doi.org/10.1093/mnras/stv1447}{\color{magenta}\mnras},
  \href{https://ui.adsabs.harvard.edu/abs/2015MNRAS.452.3529V}{452, 3529}

\bibitem[{{Wechsler} {et~al.}(2002){Wechsler}, {Bullock}, {Primack},
  {Kravtsov}, \& {Dekel}}]{wechsler02}
{Wechsler}, R.~H., {Bullock}, J.~S., {Primack}, J.~R., {Kravtsov}, A.~V., \&
  {Dekel}, A. 2002,
  \href{http://dx.doi.org/10.1086/338765}{\color{magenta}\apj},
  \href{http://adsabs.harvard.edu/abs/2002ApJ...568...52W}{568, 52}

\bibitem[{{White} \& {Rees}(1978)}]{white78}
{White}, S.~D.~M. \& {Rees}, M.~J. 1978, \mnras,
  \href{http://adsabs.harvard.edu/abs/1978MNRAS.183..341W}{183, 341}

\bibitem[{{Wu} {et~al.}(2022){Wu}, {Costanzi}, {To}, {Salcedo}, {Weinberg},
  {Annis}, {Bocquet}, {da Silva Pereira}, {DeRose}, {Esteves}, {Farahi},
  {Grandis}, {Rozo}, {Rykoff}, {Varga}, {Wechsler}, {Zeng}, {Zhang}, {Zhang},
  \& {DES Collaboration}}]{wu22}
{Wu}, H.-Y., {Costanzi}, M., {To}, C.-H., {et~al.} 2022,
  \href{http://dx.doi.org/10.1093/mnras/stac2048}{\color{magenta}\mnras},
  \href{https://ui.adsabs.harvard.edu/abs/2022MNRAS.515.4471W}{515, 4471}

\bibitem[{{Xhakaj} {et~al.}(2020){Xhakaj}, {Diemer}, {Leauthaud}, {Wasserman},
  {Huang}, {Luo}, {Adhikari}, \& {Singh}}]{xhakaj20}
{Xhakaj}, E., {Diemer}, B., {Leauthaud}, A., {et~al.} 2020,
  \href{http://dx.doi.org/10.1093/mnras/staa3046}{\color{magenta}\mnras},
  \href{https://ui.adsabs.harvard.edu/abs/2020MNRAS.499.3534X}{499, 3534}

\bibitem[{{Zhao} {et~al.}(2009){Zhao}, {Jing}, {Mo}, \& {Bn{\"o}rner}}]{zhao09}
{Zhao}, D.~H., {Jing}, Y.~P., {Mo}, H.~J., \& {Bn{\"o}rner}, G. 2009,
  \href{http://dx.doi.org/10.1088/0004-637X/707/1/354}{\color{magenta}\apj},
  \href{http://adsabs.harvard.edu/abs/2009ApJ...707..354Z}{707, 354}

\bibitem[{{Z{\"u}rcher} \& {More}(2019)}]{zurcher19}
{Z{\"u}rcher}, D. \& {More}, S. 2019,
  \href{http://dx.doi.org/10.3847/1538-4357/ab08e8}{\color{magenta}\apj},
  \href{https://ui.adsabs.harvard.edu/abs/2019ApJ...874..184Z}{874, 184}

\end{thebibliography}
\end{document}